\documentclass[journal]{IEEEtran}

\ifCLASSINFOpdf

\else

\fi

\usepackage{graphicx}
\usepackage{amsmath}
\usepackage[noend]{algpseudocode}
\usepackage{algorithmicx,algorithm}
\usepackage{amsthm}
\usepackage{amsfonts}
\usepackage{subfigure} 
\usepackage{color}
\usepackage{cite}
\usepackage{stfloats}
\usepackage{setspace}
\newtheorem{proposition}{Proposition}

\usepackage{mathrsfs,amsmath}
\usepackage{float}
\newenvironment{fequation}{\begin{equation}\footnotesize}{\end{equation}}
\newenvironment{sequation}{\begin{equation}\small}{\end{equation}} 
\makeatletter
\renewcommand{\maketag@@@}[1]{\hbox{\m@th\normalsize\normalfont#1}}%
\makeatother 
\begin{document}

	\title{On the Fundamental Trade-Offs of Time-Frequency Resource Distribution in OFDMA ISAC}

	\author{Xiao-Yang~Wang,~Shaoshi~Yang,~\IEEEmembership{Senior Member,~IEEE},~Kaitao Meng,~\IEEEmembership{Member,~IEEE}, ~Hou-Yu Zhai,~Christos~Masouros,~\IEEEmembership{Fellow,~IEEE}
		
		\thanks{X.-Y. Wang, S. Yang, H.-Y. Zhai are with the School of Information and Communication Engineering, Beijing University of Posts and Telecommunications, and the Key
			Laboratory of Universal Wireless Communications, Ministry of Education, Beijing 100876, China. X.-Y. Wang is also with the Department of Electronic and Electrical Engineering, University College London, London WC1E 7JE, UK (E-mail:  wangxy\_028@bupt.edu.cn, shaoshi.yang@bupt.edu.cn, 2hy@bupt.edu.cn).}
		
		\thanks{K. Meng and C. Masouros are with the Department of Electronic and Electrical Engineering, University College London, London WC1E 7JE, UK (E-mail: kaitao.meng@ucl.ac.uk,  c.masouros@ucl.ac.uk).}
	}

	\markboth{}%
	{Shell \MakeLowercase{\textit{et al.}}: Bare Demo of IEEEtran.cls for IEEE Journals}
	
	\maketitle
	
	\begin{abstract}
		Integrated sensing and communications (ISAC) is widely recognized as a pivotal and emerging technology for the next-generation mobile communication systems. However, how to optimize the time-frequency domain radio resource distribution for both communications and sensing, especially in scenarios where conflicting priorities emerge, becomes a crucial and challenging issue.
		In response to this problem, we first formulate the theoretical relationship between frequency domain subcarrier distribution and the range Cramér-Rao bound (CRB), and time domain sensing symbol distribution and the velocity CRB, as well as between subcarrier distribution and achievable communication rates in narrowband systems. Based on the derived range and velocity CRB expressions, the subcarrier and sensing symbol distribution schemes with the optimal and the worst sensing performance are respectively identified under both single-user equipment (single-UE) and multi-UE orthogonal frequency-division multiple access (OFDMA) ISAC systems. Furthermore, it is demonstrated that the impact of subcarrier distribution on achievable communication rates in synchronous narrowband OFDMA ISAC systems is marginal. 
		This insight reveals that the constraints associated with subcarrier distribution optimization for achievable rates can be released. 
		To substantiate our analysis, we present simulation results that demonstrate the performance advantages of the proposed distribution schemes. 
	\end{abstract}
	

	\IEEEpeerreviewmaketitle

	\section{Introduction}
	
	\IEEEPARstart{W}{ith} the increasing volume of data transfer in mobile communications, carrier frequencies are progressively shifting towards higher frequency bands, such as millimeter wave \cite{9898900} or terahertz frequency bands \cite{10061469}, owing to their abundant spectrum resources. 
	Furthermore, due to the inherent suitability of higher frequency bands for radar applications, future mobile communication systems will integrate sensing applications by leveraging unified hardware and wireless resources. This integration will result in the concept of a perceptive mobile network (PMN) as exemplified in \cite{9737357,9858656,10226306,zhang2020perceptive}. 
	
	The PMN generally operates in two modes: monostatic sensing and bistatic sensing \cite{zhang2020perceptive}. In monostatic sensing mode, the remote radio unit (RRU) performs sensing by processing echo signals transmitted by itself, while in bistatic sensing mode, the RRU performs sensing by exploiting echo signals transmitted by user equipment (UE) (namely \textit{RRU-UE bistatic sensing}) or other RRUs (namely \textit{RRU-RRU bistatic sensing}) \cite{liu2020joint}. However, achieving high-performance monostatic sensing necessitates practical full-duplex technology, which, at the time of writing, is not yet sufficiently developed for implementation due to antenna cross-talk and coupling \cite{rahman2019framework}. In contrast, since the transceivers in bistatic sensing are not co-located, there is no interference between them, and thus full-duplex technology is not required.  
	Moreover, benefiting from this distributed transceiver architecture, bistatic sensing can be conveniently performed using existing ubiquitous communication links and standards-compliant signals between RRUs and UEs \cite{zhang2020perceptive}, which does not require modifications to the current communications infrastructure and thus offers a promising low-cost sensing solution \cite{wxy,wxy-1,ni2021uplink,FarSense,zeng2020multisense}. 
	Therefore, this work investigates bistatic sensing as an example\footnote{It is important to note that the conclusions drawn in this paper are applicable to all types of OFDMA PMNs, including OFDMA monostatic sensing systems.}. 

	Bistatic sensing can be implemented using pilots, such as synchronization signal blocks (SSBs) \cite{sadiq2020techniques}, and/or data payload \cite{zhang2021enabling}. 
	Utilizing pilots offers the advantage of full compatibility with the current mobile communications technical standards \cite{wei20225g}. However, pilots only occupy a small portion of the entire frame, which may not fully exploit the potential of using communication signals for sensing. In contrast, the payload offers a much larger amount of resources for sensing than that provided by the pilots. Particularly, in payload-based sensing, data demodulated by the communication modules is treated as a specialized form of dedicated pilots and then fed to the sensing modules \cite{wxy}. The fed data is then divided by the discrete baseband orthogonal frequency-division multiplexing (OFDM) signal to extract the channel state information (CSI), a straightforward process known as data compensation specifically designed for \textit{OFDM PMNs} \cite{liu2020super}. Furthermore, accurate sensing can be implemented utilizing the large amount of extracted CSI. 
	
	In addition to the data volume, the time-frequency domain distribution of the sensing resource also affects the sensing performance of OFDM PMNs.
	Specifically, the authors of \cite{10288116} optimize the distribution of the demodulation reference signal (DMRS) in the frequency domain for narrowband systems. 
	It is proven that the optimization improves the delay estimation performance. 
	Moreover, the authors of \cite{8628347} also optimize the frequency domain pilot distribution in narrowband systems and design a new pilot pattern called the stepped pattern. This pilot pattern extends the maximum perceptible range while maintaining a constant range estimation Cramér-Rao bound (CRB). Additionally, this design is further validated by a hardware implementation in \cite{9062788}. 
	
	
	
	Despite the above work \cite{10288116,8628347,9062788}, the theoretical relationship between time-frequency domain resource distribution and range/velocity estimation CRB has not been comprehensively investigated yet. 
	Concretely, the DMRS distribution in \cite{10288116} is investigated only for improving range estimation CRB, solely under resource-block-group-level Type-0 DMRS resource distribution, one of three DMRS distribution schemes in the fifth-generation technology standard for cellular networks (5G) \cite{3gpp.38.214}. Moreover, pilot distribution in \cite{8628347} is only improved to increase the maximum perceptible range, not to improve the range CRB. Additionally, both studies do not consider optimization for the velocity CRB and do not delve into the trade-off analysis between resource distribution and achievable communication rates. Hence, the conclusions and insights provided by these works do not provide a fundamental and general enough framework to guide time-frequency domain resource distribution design for single-UE orthogonal frequency-division multiple access (OFDMA) PMNs (namely OFDM PMNs), let alone for \textit{multi-UE multistatic OFDMA PMNs}\footnote{In this paper, time-frequency domain resource distribution refers to the joint distribution of time domain sensing symbols and frequency domain subcarriers.}. 

Optimization of resource distribution for OFDMA PMNs is much more critical than for OFDM PMNs, due to the more widespread use of OFDMA and its greater sensing  potential in cellular communication systems \cite{4287203,10091198,meng2024cooperative}.  
Specifically, OFDMA could facilitate simultaneous and independent communication links between a single RRU and multiple UEs (or RRUs), enabling multistatic sensing and communications (ISAC) in cellular systems. Multistatic ISAC systems generally offer enhanced sensing capabilities due to their richer frequency and spatial resources \cite{tagliaferri2023cooperative}. This supports more sophisticated applications, such as multistatic environmental reconstruction and imaging \cite{9534682,10404430,10097213}, and multistatic vehicle sensing \cite{tagliaferri2023cooperative}, which require enhanced sensing capabilities. However, there is no research on time-frequency domain resource distribution for OFDMA PMNs. 
Therefore, optimizing resource distribution within OFDMA PMNs is indispensable and pressing. 

It is much more challenging to optimize the resource distribution for multi-UE OFDMA PMNs than for single-UE OFDMA PMNs.
Let us first take frequency domain subcarrier distribution optimization in OFDM PMN scenarios as an example. The optimization mainly involves selecting a certain number of integer indexes of subcarriers from a given set for pilot insertion to enhance communications and sensing performance of a single UE, which is an integer optimization problem. For frequency domain subcarrier distribution optimization in OFDMA PMNs, the subcarrier indexes are partitioned into multiple subsets, each assigned to a different UE. The partitioning aims not only to enhance the sensing or communications performance of each UE but also to balance the performance trade-offs across multiple RRU-UE links. This is the non-deterministic polynomial-time hard (NP-hard) cardinality-constrained  multi-way number partitioning (MWNP) problem \cite{SIAM000}. Although some optimization algorithms have been proposed for the MWNP problem, they are designed for specific optimization objectives and constraints, such as the scheme for minimizing the largest sum with cardinality constraints \cite{MathematicalMethodsofOperationsResearch} and 
the scheme \cite{JournalofScheduling} for the more general objectives, namely optimizing functions of the sum of the subsets, without cardinality constraints. 
However, these objectives and constraints differ from the optimization objective in this paper, i.e. maximizing the minimum variance of the subsets with constraints on the number of assigned subcarriers. Therefore, an optimization scheme for this MWNP problem with constraints on the variance and cardinality of each subset needs to be designed.


Motivated by the discussion above, our objectives are: 1) to implement payload-based OFDMA multistatic sensing; 2) to comprehensively investigate the quantitative relationship between the distribution of time-frequency domain sensing resources and both the sensing CRB and achievable communication rates; 3) to optimize the time-frequency domain sensing resource distribution for both single-UE and multi-UE OFDMA PMNs. 
Specifically, the key contributions are summarized below.
\begin{itemize}
	\item We propose the concept of \textit{OFDMA multistatic sensing}. To implement this sensing using data payload, we introduce a novel compensation matrix design framework to obtain interference-free CSI for each UE in payload-based \textit{OFDMA PMNs} and outline the necessary prerequisites. This framework provides a new perspective on acquiring interference-free CSI in payload-based OFDMA PMNs.
	
	\item We derive the joint range and velocity estimation CRB in OFDMA PMNs, revealing that the range and velocity estimation CRB for each UE are inversely proportional to the variance of the subcarrier indexes and sensing symbol indexes, respectively. 
	
	\item In single-UE OFDMA PMNs, we formulate an integer optimization problem to optimize the sensing CRB. It is proved that the edge-first and subband distribution result in the best and worst CRB for both range and velocity estimation, respectively. This conclusion facilitates achieving optimal sensing performance within the given resources. 
	
	\item In multi-UE OFDMA PMNs, we formulate a cardinality-constrained MWNP problem to minimize the maximum sensing CRB of multiple UEs and propose a tight CRB lower bound for the problem. Moreover, we find that the CRB of the interleaved distribution scheme almost reaches the lower bound. This suggests that the interleaved distribution scheme is sufficiently effective in practice and does not justify the excessively high computational complexity of finding the exact optimal solution. 
	
	\item In OFDMA PMNs, we observe that subcarrier distribution has a minimal impact on achievable communication rates in narrowband synchronous systems. By formulating the achievable communication rates, we ascertain that subcarrier distribution schemes influence these rates by affecting inter-channel interference (ICI). However, irrespective of the subcarrier distribution scheme employed, the power of the ICI remains negligible. This finding underscores the minimal impact of subcarrier distribution on achievable rates.
\end{itemize}

Our analysis and simulation results indicate that the edge-first distribution achieves optimal sensing performance in single-UE OFDMA PMNs, while the interleaved distribution approaches optimal sensing performance in multi-UE OFDMA PMNs. Conversely, the subband distribution consistently yields the poorest sensing performance in both single-UE and multi-UE OFDMA PMNs. Furthermore, the negligible impact of ICI indicates that the subcarrier distribution minimally affects achievable communication rates.

The structure of this paper is outlined as follows. In Section II, we present the system model for OFDMA PMNs. In Section III, we describe the implementation of multistatic range and velocity sensing in payload-based OFDMA PMNs. In Section IV, we derive the CRB for range and velocity sensing and optimize the subcarrier and sensing symbol distribution in both single-UE and multi-UE OFDMA PMNs. Moreover, in Section V, we analyse the impact of the subcarrier distribution on the achievable rates. Section VI presents numerical simulations to verify our analysis. Finally, our conclusions are presented in Section VII.

\textit{Notations}: ${\bf A}^{\textrm{T}},{\bf A}^{*}, {\bf A}^{\textrm{H}}$, and ${\bf A}^{-1}$ represent the transpose, conjugate, conjugate transpose, and inverse of the matrix ${\bf A}$, respectively; ${\textrm{diag}}({\bf a}_1,\cdots,{\bf a}_n)$ is a block diagonal matrix with the diagonal blocks $\{{\bf a}_1,\cdots,{\bf a}_n\}$. ${\bf a}[m]$ and ${\bf A}[m,n]$ are the $m$th element of the vector ${\bf a}$ and the $(m,n)$th element of ${\bf A}$, respectively; ${a}(t)$, ${\bf a}(t)$, and ${\bf A}(t)$ are the scalar function, the vector function, and the matrix function with respect to $t$, respectively; ${\bf I}_N$ and ${\bf 0}_{M\times N}$ are the $N\times N$ identity matrix and the $M\times N$ all-zero matrix, respectively; $j$, $(\star)$, $\otimes$, $\operatorname{Re}(\cdot)$, ${\operatorname{Im}(\cdot)}$, and ${\operatorname{E}}(\cdot)$ are defined as the imaginary unit, the convolution operator, the Kronecker product operator, the real part taking operator, the imaginary part taking operator, and the expectation operator, respectively. 
Moreover, $\operatorname{card}\{\cdot\}$ represents the number of elements in a set, and $\mathbb{Z}$ is defined as the integer set. Finally, $\operatorname{Vec}(\cdot)$, $\operatorname{det}(\cdot)$ and $\operatorname{log}(\cdot)$ are the vectorization operator, the determinant operator, and the logarithm operator, respectively.

\section{System Model}
\begin{figure}[tbp]
	\centering
	\includegraphics[width=3.2in]{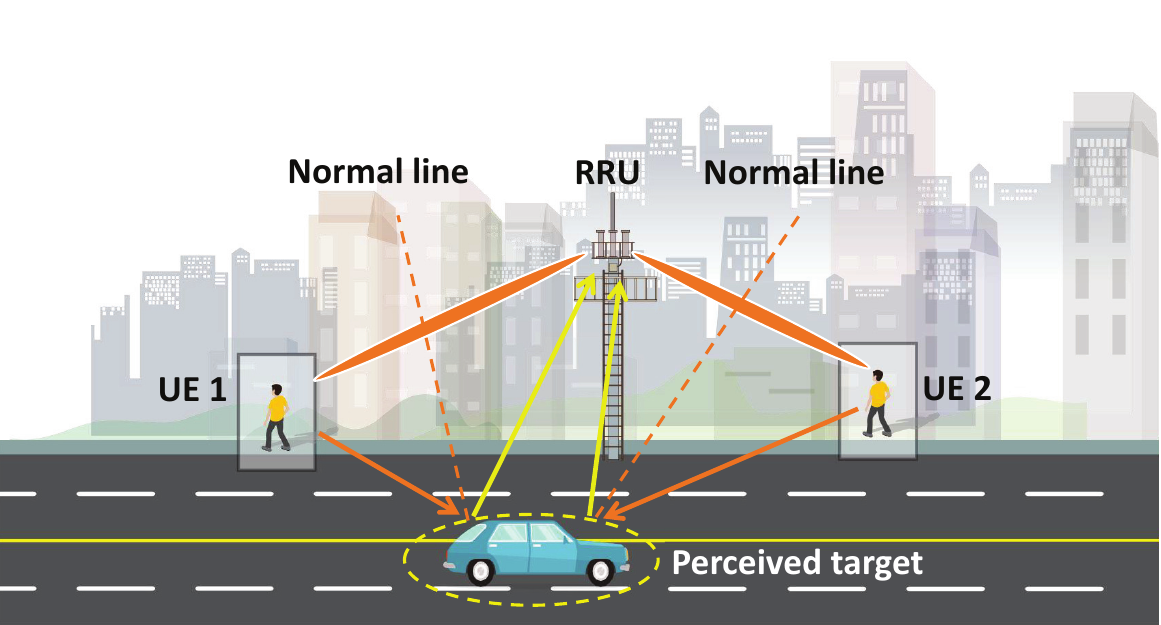}
	\caption{System model.}\label{Sketch}
	\vspace*{-3pt}
\end{figure}
In this work, we consider a multiple-input multiple-output (MIMO)-OFDMA PMN, as depicted in Fig. \ref{Sketch}. The $M_{\textrm{R}}$-antenna RRU receives signals transmitted by $K$ $M_{\textrm{T}}$-antenna UEs. The propagation delay of the $l$th path from the $k$th UE to the target and then to the receiver, as well as the velocity of the reflector projected on the normal direction in this path, are denoted as $\tau_{k,l}$ and $v_{k,l}$, respectively. The RRU will perform range and velocity sensing utilizing the received communication signals.  
Following the assumption in \cite{rahman2019framework,zhang2021enabling}, the entire data payload is exploited to perform the sensing. Note that the signal processing aspect in this scenario is the same as that in pilot-based sensing.

Let us denote the carrier frequency, the total number of subcarriers, and the subcarrier spacing as $f_\textrm{c}$, $N$, and $\Delta f$, respectively. Furthermore, $N_k$ subcarriers are assigned to the $k$th UE, and the data modulated on this $N_k$ subcarriers in the $g$th OFDM symbol is denoted as ${\bf s}_{g,k}\in\mathbb{C}^{1\times N_k}$. Then, we define ${\boldsymbol \zeta}_k\in\mathbb{C}^{1\times N_k}$ as the subcarrier distribution vector of the $k$th UE, where ${\boldsymbol \zeta}_k[n]$ is the index of the $n$th subcarrier of the $k$th UE. 

Therefore, the analog signal transmitted by the $k$th UE is
\begin{equation}\label{x_k(t)}
	{\bf x}_{g,k}(t)=e^{j2\pi f_{\textrm{c}}t}{\boldsymbol \varpi}_k\sum\nolimits_{n=1}^{N_k}{\bf s}_{g,k}[n]e^{j2\pi \boldsymbol{\zeta}_k[n]\Delta ft}.
\end{equation}where ${\boldsymbol \varpi}_k\in\mathbb{C}^{M_{\textrm{T}}\times 1}$ represents the beamforming vector of the $k$th UE. We also denote the time-varying delay function for the $l$th path of the $k$th UE with respect to time instant $t$ as $\tau_{k,l}^\textrm{F}(t)=\frac{2v_{k,l}}{c}t+\tau_{k,l}$, where $c$ is the speed of light. Then, the time-varying channel impulse response function of the $l$th path of the $k$th UE can be formulated as
\begin{equation}\label{h_kl^n(t)}
	{\bf H}_{k,l}(t,\tau_{k,l}^\textrm{F}(t))=\alpha_{k,l}\delta(t-\tau_{k,l}^\textrm{F}(t)){\bf a}(\Omega_{k,l}^{\textrm{r}}){\bf a}^{\textrm{T}}(\Omega_{k,l}^{\textrm{t}}),
\end{equation}
where $\alpha_{k,l}$ is the channel gain of the $l$th path of the $k$th UE. Since the bandwidth of the system is much smaller than the carrier frequency, we assume that the system is narrowband, as many other studies do \cite{10288116,8628347,9062788,wei20225g,wxy,ni2021uplink}. Therefore, $\alpha_{k,l}$ is a constant for different subcarriers for the $l$th path of the $k$th UE. Moreover, $\delta(t)$ is the impulse function. ${\bf a}(\Omega_{k,l}^{\textrm{r}})=[1,e^{j\Omega_{k,l}^{\textrm{r}}},\cdots,e^{j(M_{\textrm{R}}-1)\Omega_{k,l}^{\textrm{r}}}]^{\textrm{T}}$ and ${\mathbf{a}}(\Omega_{k,l}^{\textrm {t}})=[1,e^{j\Omega_{k,l}^{\textrm{ }}},\cdots,e^{j(M_{\textrm{T}}-1)\Omega_{k,l}^{\textrm{t}}}]^{\textrm{T}}$ are the receiving and the transmitting steering vectors of the signal from the $k$th UE along its $l$th path, respectively. Furthermore, $\Omega_{k,l}^{\textrm{r}}=2\pi d\cos(\theta_{k,l}^{\textrm{r}})/\lambda$, $\Omega_{k,l}^{\textrm{t}}=2\pi d\cos(\theta_{k,l}^{\textrm{t}})/\lambda$, $d$ is the antenna spacing, and $\lambda$ is the wavelength of the signal. Additionally, $\theta_{k,l}^{\textrm{r}}$ and $\theta_{k,l}^{\textrm{t}}$ denote the angle of arrival (AOA) and the angle of departure (AOD) of the signal from the $k$th UE along its $l$th propagation path, respectively. The received equivalent baseband signal can then be expressed as
\begin{fequation}\label{y_g^n(t)}
	\begin{aligned}
		{\bf y}_g(t)&=e^{-j2\pi f_{\textrm{c}}t}\sum\nolimits_{k=1}^{K}\sum\nolimits_{l=1}^{L_k} [{\bf H}_{k,l}(t)\star{\bf x}_{g,k}(t)]+{\bf w}_g(t)\\
		&=\! \sum\nolimits_{k=1}^{K}\!\sum\nolimits_{l=1}^{L_k}\!\!\alpha_{k,l}{\bf a}(\Omega_{k,l}^{\textrm{r}}){\bf a}^{\textrm{T}}(\Omega_{k,l}^{\textrm{t}}){\boldsymbol \varpi}_k\!\sum\nolimits_{n=1}^{N_k}\!{\bf s}_{g,k}[n]e^{-j2\pi f_{\textrm{c}}\tau_{k,l}}\\
		&\cdot e^{-j2\pi [f_{\textrm{c}}+{\boldsymbol{\zeta}}_k[n]\Delta f]\frac{2v_{k,l}}{c}(t-\tau_{k,l})} e^{j2\pi \boldsymbol{\zeta}_k[n]\Delta f(t-\tau_{k,l})}+{\bf w}_g(t),
	\end{aligned}
\end{fequation}%
where $L_k$ and ${\bf w}_g(t)$ represent the number of paths of the $k$th UE and the complex additive white Gaussian noise (AWGN) vector function with zero mean and covariance matrix $\sigma^2{\bf I}_{M_{\textrm{R}}}$, respectively. For (\ref{y_g^n(t)}), it is worth noting that the term ${\boldsymbol{\zeta}}_k[n]\Delta f\frac{2v_{k,l}}{c}t$ is small on the time scale of hundreds of OFDM symbols and $\frac{2 v_{k,l}f_\textrm{c}}{c}t$ is small enough on the time scale of a single OFDM symbol. Thus, as described in \cite{liu2020super}, we approximate $ e^{-j2\pi{\boldsymbol{\zeta}}_k[n]\Delta f\frac{2v_{k,l}}{c}t}$ and $e^{-j2\pi\frac{2 v_{k,l}f_\textrm{c}}{c}t}$ as constants within hundreds of symbols and a single symbol, respectively.

Furthermore, the $g$th discrete OFDM symbol ${\bf Y}_g\in\mathbb{C}^{M_\textrm{R}\times N}$ will be sampled from ${\bf y}_g(t)$. Let us assume that the length of an OFDM sample is $T_{\textrm {sam}}$, the length of the cyclic prefix is $T_{\textrm{c}}$, and the length of an entire OFDM symbol is $T_{\textrm{s}}$. Then, ${\bf Y}_g=$ $[{\bf y}_g(T_\textrm{sam}),\!\cdots\!,{\bf y}_g(NT_\textrm{sam})]$ can be approximately formulated as
\begin{equation}\label{Y_g}
	\begin{aligned}
		{\bf Y}_g
		\!\approx&\sum\nolimits_{k=1}^{K}\sum\nolimits_{l=1}^{L_k}e^{-j2\pi f_{\textrm{c}}\{\frac{2v_{k,l}}{c}[({\boldsymbol{\psi}}_k[g]-1)T_{\textrm{s}}+T_{\textrm{c}}]\}}\\
		&\cdot\bar\alpha_{k,l}{\bf a}(\Omega_{k,l}^{\textrm{r}}){\bf a}^{\textrm{T}}(\Omega_{k,l}^{\textrm{t}}){\boldsymbol \varpi}_k{\boldsymbol\tau}_{k,l}{\bf \Gamma}_k{\bf D}_{g,k}{\bf F}_N\!+\!{\bf W}_g,
	\end{aligned}
\end{equation}
where  ${\boldsymbol\tau}_{k,l}=[e^{-j2\pi{\boldsymbol{\zeta}}_k[1]\Delta f\tau_{k,l}},\cdots,e^{-j2\pi {\boldsymbol{\zeta}}_k[N_k]\Delta f\tau_{k,l}}]$ and $\bar\alpha_{k,l}=$ $\alpha_{k,l}e^{-j2\pi f_{\textrm{c}}\tau_{k,l}(1-\frac{2v_{k,l}}{c})}$, respectively. ${\bf D}_{g,k}$ is a diagonal matrix whose diagonal vector is ${\bf s}_{g,k}{\bf \Gamma}_k\in\mathbb{C}^{1\times N}$. ${\bf \Gamma}_k\in\mathbb{Z}^{N_k\times N}$ is the subcarrier distribution matrix for the $k$th UE. Specifically, ${\bf \Gamma}_k[i,{\boldsymbol \zeta}_k[i]]$, $i=1,\cdots,N_k$, is 1, while the other elements are 0. Additionally,  ${\bf F}_N$ represents the $N$-dimensional inverse fast Fourier transform (IFFT) matrix, and ${\bf W}_g=[{\bf w}_g(T_\textrm{sam}),\cdots,{\bf w}_g(NT_\textrm{sam})]$. Moreover, $\boldsymbol{\psi}_k\in\mathbb{Z}^{1\times G_k}$ denotes the index of the $G_k$ sensing symbols assigned to the $k$th UE, where $G_k\le G$. This limitation arises because the bandwidth between the communication module and the sensing module may not support feeding a large amount of the payload. Therefore, we assume that only a portion of the payload, which is still significantly larger than the number of pilots, is fed from the communication module to the sensing module. In a limiting scenario, $G_k$ can also be set to $G$.

\section{Target Sensing in OFDMA PMNs}

This section aims to achieve multistatic sensing in OFDMA PMNs and is divided into two subsections. The first subsection acquires CSI by compensating data payload with designated compensation matrices. Additionally, the requirements concerning subcarrier distribution for successful compensation are also investigated. For brevity, here we assume that the data payload has been successfully detected and fed to the sensing module. 
In the second subsection, a two-dimensional multiple signal classification (2-D MUSIC) algorithm is applied to estimate the ranges and velocities of targets, serving as a sensing example in OFDMA PMNs. 

\subsection{Data Compensation for OFDMA PMNs}
In this subsection, we design compensation matrices to remove  data information from the payload and acquire the CSI needed for sensing. Specifically, the data compensation process is depicted in \cite{liu2020super} as follows: 1) the communication module of the RRU first feeds the demodulated data to the sensing module; 2) the fed data is then simply divided by the discrete baseband OFDM signals in the sensing module to obtain the CSI.

It should be noted that the compensation method in \cite{liu2020super} is specifically designed for OFDM PMNs. This method is primarily based on empirical observations and lacks a comprehensive design framework. 
Therefore, we investigate the data compensation in OFDMA PMNs from an interference elimination perspective, aiming to obtain the CSI of each UE without interference from the other UEs. 

First, we convert ${\bf Y}_g$ to the frequency domain as
\begin{equation}\begin{small}
			\label{compensation1}
			\begin{aligned}
				{\bf Y}_g{\bf F}_N^{-1}=&\sum\nolimits_{k=1}^{K}\sum\nolimits_{l=1}^{L_k}e^{-j2\pi f_{\textrm{c}}\{\frac{2v_{k,l}}{c}[(\boldsymbol{\psi}_k[g]-1)T_\textrm{s}+T_\textrm{c}]\}}\\
				&\cdot\bar\alpha_{k,l}{\bf a}(\Omega_{k,l}^{\textrm{r}}){\bf a}^{\textrm{T}}(\Omega_{k,l}^{\textrm{t}}){\boldsymbol \varpi}_k{\boldsymbol\tau}_{k,l}{\bf \Gamma}_k{\bf D}_{g,k}\!+\!{\bf W}_g{\bf F}_N^{-1}.
			\end{aligned}
		\end{small}
\end{equation}
We then denote $K$ desired compensation matrices for each UE as ${\bf C}_k\in\mathbb{C}^{N\times N}$, where $k=1,\cdots,K$. To acquire CSI of each UE without interference caused by the compensation, we establish the following equation that the compensation matrices ${\bf C}_k, k=1,\cdots,K$ should satisfy
\begin{equation}\label{s.t.}
		\begin{small}
			\begin{aligned}
				{\bf Y}_g{\bf F}_N^{-1}{\bf C}_k\!=&\!\Big[\sum\nolimits_{l=1}^{L_k}e^{-j2\pi f_\textrm{c}\{\frac{2v_{k,l}}{c}[(\boldsymbol{\psi}_k[g]-1)T_\textrm{s}+T_\textrm{c}]\}}\bar\alpha_{k,l}\\
				&\cdot{\bf a}(\Omega_{k,l}^{\textrm{r}}){\bf a}^{\textrm{T}}(\Omega_{k,l}^{\textrm{t}}){\boldsymbol \varpi}_k{\boldsymbol\tau}_{k,l}\Big]{\boldsymbol{\Gamma}_k}+{\bf W}_g{\bf F}_N^{-1}{\bf C}_k.
			\end{aligned}
		\end{small}
\end{equation}
To find solutions, we split (\ref{s.t.}) into separate equations as
\begin{equation}\label{condition}
		\begin{small}
			\left\{
			\begin{aligned}
				&{\bf D}_{g,k}{\bf C}_k={\bf \Gamma}_k^\textrm{T}{\bf \Gamma}_k,\\
				&\Big[\sum\nolimits_{k^{'}\neq k}^{K}\sum\nolimits_{l=1}^{L}\tilde\alpha_{k^{'},l}{\bf a}(\Omega_{k^{'},l}^{\textrm{r}}){\boldsymbol\tau}_{k^{'},l}{\bf D}_{g,k^{'}}\Big]{\bf C}_{k} = {\bf 0}_{M_{\textrm{R}}\times N},
			\end{aligned}
			\right.
		\end{small}
\end{equation}
where $\tilde\alpha_{k^{'},l}=e^{-j2\pi f_\textrm{c}\{\frac{2v_{k,l}}{c}[(\boldsymbol{\psi}_k[g]-1)T_\textrm{s}+T_\textrm{c}]\}}\bar\alpha_{k,l}{\bf a}^{\textrm{T}}(\Omega_{k,l}^{\textrm{t}}){\boldsymbol\varpi}_k$. Particularly, the first equation aims to completely remove the information of the $g$th symbol of the $k$th UE, while the second equation completely eliminates the interference.

Moreover, it is important to note that if interference can be tolerated, the second equation can be modified to $ [\sum_{k^{'}\neq k}^{K}\sum_{l=1}^{L}\tilde\alpha_{k^{'},l}{\bf a}(\Omega_{k^{'},l}^{\textrm {r}}){\boldsymbol\tau}_{k^{'},l}{\bf D}_{g,k^{'}}]{\bf C}_{k}\preceq{\bf P}$, where $\!\preceq\!$ is defined as a matrix relationship where each element of the left matrix is smaller than or equal to the corresponding element of the right matrix. Moreover, ${\bf P}[m,n]$ is the upper limit of acceptable interference power at the $n$th subcarrier of signals received by the $m$th antenna. This modification will extend the scope of the feasible compensation matrices and may loosen the prerequisites for the existence of a feasible solution. However, this is not the focus of this paper. Therefore, for simplicity, we set ${\bf P}={\bf 0}_{M_{\textrm{R}}\times N}$ in the following. 

Then, concerning (\ref{condition}), Proposition 1 below outlines the solutions of (\ref{s.t.}), namely ${\bf C}_k,$ $k=1,\cdots,K$, and the condition under which solutions ${\bf C}_k$ exist.
\begin{proposition}
		For $k, k^{'}=1,\!\cdots\!,K$ and $k\neq k^{'}$, ${\bf C}_k$ satisfying (\ref{condition}) exists, if and only if, for $g\!=\!1,\!\cdots\!, G$, ${\bf D}_{g,k}$ satisfies ${\bf D}_{g,k}{\bf D}_{g,k^{'}}={\bf 0}_{N\times N}$ and {\rm${\bf C}_k={\textrm{diag}}({\bf c}_k)$}, where ${\bf c}_k$ is
		\begin{equation}
			{\bf c}_k[n]=
			\begin{cases}
				({\bf D}_{g,k}[n,n])^{-1},&\ {\bf D}_{g,k}[n,n]\neq 0,\\
				0,&\ {\bf D}_{g,k}[n,n]= 0,\\
			\end{cases}
		\end{equation}
\end{proposition}
\textit{Proof}: See Appendix A.

After the compensation, range and velocity sensing can be further implemented in the following section. 

\subsection{Range and Velocity Sensing in OFDMA PMNs}
In this subsection, the 2-D MUSIC algorithm is applied as an example to estimate range-velocity pairs using the CSI obtained in subsection A. Nevertheless, our framework can be generalized to arbitrary receivers through the analysis of the CRB in the following section. 

First, let us define the $m$th row of ${\bf Y}_g{\bf F}_N^{-1}{\bf C}_{k}$ as ${{\bf y}}_{m,k,g}$, and stack ${{\bf y}}_{m,k,g}$ as matrix $\tilde{{\bf Y}}_{m,k}\in\mathbb{C}^{G_k\times N}=[{{{\bf y}}_{m,k,1}}^{\textrm {T}},\cdots,{{{\bf y}}_{m,k,G_k}}^{\textrm{T}}]^{\textrm{T}}$. Since only $N_k$ subcarriers bear data for the $k$th UE, there are $N-N_k$ all-zero columns in $\tilde{{\bf Y}}_{m,k}$. Thus, we neglect these all-zero columns and obtain $\tilde{{\bf Y}}_{m,k}\boldsymbol{\Gamma}^{\textrm{T}}$, which essentially represents the data demodulated from the signals of the $k$th UE and received by the $m$th antenna of the RRU. Additionally, $\tilde{{\bf Y}}_{m,k}\boldsymbol{\Gamma}^{\textrm {T}}$ can be vectorized as $\bar{{\bf y}}_{m,k}\in\mathbb{C}^{G_kN_k\times 1}$:
\begin{sequation}\label{ymk}
		\begin{aligned}
			\bar{{\bf y}}_{m,k}&=\textrm{Vec}(\tilde{{\bf Y}}_{m,k}\boldsymbol{\Gamma}^{\textrm {T}})=\sum\nolimits_{l=1}^{L_k}\beta_{m,k,l}({\boldsymbol\xi}_{k,l}\otimes{{\boldsymbol\tau}}_{k,l}^{\textrm{T}})+\bar{\bf w}_{m,k}\\
			&={\bf A}{\bf e}_k+\bar{\bf w}_{m,k},
		\end{aligned}
\end{sequation}where $\beta_{m,k,l} =\bar\alpha_{k,l}e^{-j2\pi f_\textrm{c}\frac{2v_{k,l}}{c}T_\textrm{c}}e^{jm\Omega_{k,l}^{\textrm{r}}}{\bf a}^{\textrm {T}}(\Omega_{k,l}^{\textrm{t}}){\boldsymbol \varpi}_k$, ${\boldsymbol\xi}_{k,l}= [1,\!\cdots\!, e^{-j2\pi f_\textrm{c}(\boldsymbol{\psi}_k[G_k]-1)T_\textrm{s}\frac{2v_{k,l}}{c}}]^{\textrm {T}}$, and ${\bf e}_k=$ $[\beta_{m,k,1},\cdots,\beta_{m,k,L_k}]^{\textrm{T}}$. Moreover, ${\bf A}=[{\boldsymbol\xi}_{k,1}\otimes{{\boldsymbol\tau}}_{k,1}^{\textrm {T}},\cdots,$ ${\boldsymbol\xi}_{k,L_k}\otimes{{\boldsymbol\tau}}_{k,L_k}^{\textrm{T}}]$, and $\bar{\bf w}_{m,k}=$ $\operatorname{Vec}({\bf W}_g{\bf F}_N^{-1}{\bf C}_k\boldsymbol{\Gamma}^{\textrm {T}})$.

The covariance matrix of $\bar{{\bf y}}_{m,k}$ is then computed as
\begin{equation}\label{equ13}
		\begin{aligned}
			{\bf R}_{f,\tau}&\!=\operatorname{E}(\bar{{\bf y}}_{m,k}(\bar{{\bf y}}_{m,k})^{\textrm {H}})={\bf A}{\bf R}_{b}{\bf A}^{\textrm{H}}+\bar{\bf w}_{m,k}\bar{\bf w}_{m,k}^{\textrm{H}},
		\end{aligned}
\end{equation}
where ${\bf R}_b={\bf e}_k{\bf e}_k^{\textrm{H}}$.
Moreover, by applying the eigenvalue decomposition to ${\bf R}_{f,\tau}$, we obtain
\begin{equation}
		\begin{aligned}
			{\bf R}_{f,\tau}=[{\bf U}_{{\textrm{s}}},{\bf U}_{{\textrm {n}}}]{\boldsymbol{\Lambda}}_{f,\tau}[{\bf U}_{{\textrm{s}}},{\bf U}_{{\textrm {n}}}]^{\textrm{H}}.
		\end{aligned}
\end{equation}
where ${\bf U}_{{\textrm {s}}}$ and ${\bf U}_{{\textrm {n}}}$ are the normalized unitary matrices, and ${\boldsymbol{\Lambda}}_{f,\tau}$ is the diagonal matrix whose diagonal elements are the eigenvalues of ${\bf R}_{f,\tau}$. Consequently, time delays and Doppler offsets can be estimated by finding the coordinates of the peaks of the metric
\begin{equation}
		\begin{aligned}
			P(f,\tau)&={1}/{[({\boldsymbol\xi}_{k,l}\otimes{\boldsymbol\tau}_{k,l})^{\textrm {H}}{\bf U}_{{\textrm {n}}}{\bf U}_{{\textrm {n}}}^{\textrm {H}}({\boldsymbol\xi}_{k,l}\otimes{\boldsymbol\tau}_{k,l})]}.
		\end{aligned}
\end{equation}
By adopting this estimation scheme for $k=1,\cdots,K$, the range and velocity parameter pairs corresponding to all of the $K$ UEs can be estimated, respectively. In the next section, the CRB for range and velocity will be derived based on the notations provided in this section.

\section{Range and Velocity Sensing CRB Optimization}
To characterize the mathematical relationships between the range estimation CRB and the subcarrier distribution, as well as the velocity estimation CRB and the sensing symbol distribution, the CRB for range and velocity are formulated in this section. We then optimize the CRB with respect to the subcarrier and sensing symbol distribution vector, $\boldsymbol{\zeta}_k$ and $\boldsymbol{\psi}_k$, of all UEs under both single-UE and multi-UE OFDMA PMNs.  

\subsection{Range and Velocity Sensing CRB Derivations}
In this subsection, we derive the CRB for range and velocity under unknown parameters:  $\{\beta_{m,k,1}, \cdots, \beta_{m,k,L_k}\}_{k=1}^{K}$, $\{2v_{k,1}/{c},\cdots,2v_{k,L_k}/{c} \}_{k=1}^{K}$ and $\{\tau_{k,1},\cdots,\tau_{k,L_k}\}_{k=1}^{K}$. 

\vspace{3pt}
For brevity,  we represent ${\boldsymbol \beta}_{m,k}$ and ${\bf p}_k$ respectively as
\begin{sequation}
			{\boldsymbol \beta}_{m,k}\!=\!\![{\operatorname{Re}}(\beta_{m,k,1}),{\operatorname{Im}}(\beta_{m,k,1}),\!\cdots,\!{\operatorname{Re}}(\beta_{m,k,L_k}),{\operatorname{Im}}(\beta_{m,k,L_k})],
\end{sequation}
\begin{equation}
	{\bf p}_k=[\tau_{k,1},\cdots,\tau_{k,L_k},{2v_{k,1}}/{c},\cdots,{2v_{k,L_k}}/{c}].
\end{equation}
Then, the likelihood function $f(\bar{{\bf y}}_{m,k}|{\boldsymbol \beta}_{m,k},{\bf p}_k)$ can be formulated as
\begin{equation}
	\begin{small}
		\begin{aligned}
			&f(\bar{{\bf y}}_{m,k}|{\boldsymbol \beta}_{m,k},{\bf p}_k)={1}/{\left(2\pi(\sigma_{m,k})^2\right)^{{G_kN_k}/2}}\\
			&\cdot{\textrm{exp}}\Big({\!-\!\frac{1}{2 (\sigma_{m,k})^2}\! \sum\nolimits_{g=1}^{G_k}\sum\nolimits_{n=1}^{N_k}}\Big|{\bf y}_{m,k,g}[n]-{\bf z}_{m,k,g}[n]\Big|^2\Big),
		\end{aligned}
	\end{small}
\end{equation}%
where ${\bf z}_{m,k,g}[n]$ $=\sum\nolimits_{l=1}^{L}\beta_{m,k,l}e^{-j2\pi f_\textrm{c}(\boldsymbol{\psi}_k[g]-1)T_\textrm{s}\frac{2v_{k,l}}{c}}$ $e^{-j2\pi {\boldsymbol{\zeta}}_k[n]\Delta f\tau_{k,l}}$.
Furthermore, the logarithm of the likelihood function, $L(\bar{{\bf y}}_{m,k}|{\boldsymbol \beta}_{m,k},{\bf p}_k)$, can be expressed as
\begin{equation}
	\begin{small}
		\begin{aligned}
			&L(\bar{{\bf y}}_{m,k}|{\boldsymbol \beta}_{m,k},{\bf p}_k)=-({G_kN_k}/{2}) \ln [2\pi (\sigma_{m,k})^2]-\\
			&{1}/{[2(\sigma_{m,k})^2]}\sum\nolimits_{g=1}^{ G_k}\sum\nolimits_{n=1}^{ N_k}\left|{\bf y}_{m,k,g}[n]-{\bf z}_{m,k,g}[n]\right|^2.
		\end{aligned}
	\end{small}
\end{equation} 
Utilizing $L(\bar{{\bf y}}_{m,k}|{\boldsymbol \beta}_{m,k},{\bf p}_k)$, the Fisher matrix, ${\bf J}$, can be expressed \begin{equation}\label{Fisher}
		\begin{small}
			{\bf J} =\left[
			\begin{matrix}
				{\mathrm E}(\frac{\partial^2 L}{\partial {\boldsymbol \beta}_{m,k}\partial ({\boldsymbol \beta}_{m,k})^{\textrm {T}}}) & {\mathrm E}(\frac{\partial^2 L}{\partial {\boldsymbol \beta}_{m,k}\partial {\bf p}_k^{\textrm{T}}})\\
				{\mathrm E}(\frac{\partial^2 L}{\partial {\bf p}_k\partial ({\boldsymbol \beta}_{m,k})^{\textrm{T}}}) & {\mathrm E}(\frac{\partial^2 L}{\partial {\bf p}_k\partial {\bf p}_k^{\textrm {T}}})
			\end{matrix}\right].
		\end{small}
\end{equation}
The CRB for ${\bf p}_k$, namely ${\textrm{CRB}}({\bf p}_k)$, can then be formulated as (refer to \textit{Theorem 4.1} of \cite{stoica1990performance})
\begin{sequation}\label{equ19}
		\begin{aligned}
			{\textrm{CRB}}^{-1}({\bf p}_k)\!=\!\frac{2}{(\sigma_{m,k})^2}\! \operatorname{Re}\left\{\!{\bf E}_{k}^* {\bf V}^*\!\left[{\bf I}\!-\!{\bf A}\left({\bf A}^* {\bf A}\right)^{-1}\!{\bf A}^*\right]\!{\bf V} {\bf E}_{k}\!\right\},
		\end{aligned}
\end{sequation}where ${\bf E}_{k}=\textrm{diag}([{\bf e}_{k}^{\textrm{T}}\ {\bf e}_{k}^{\textrm{T}}])$ and ${\bf V}$ is defined as $[{\partial({\boldsymbol\xi}_{k,1}\otimes{{\boldsymbol\tau}}_{k,1}^{\textrm {T}})}/{\partial \tau_{k,1}},\cdots,{\partial({\boldsymbol\xi}_{k,L_k}\otimes{{\boldsymbol\tau}}_{k,L_k}^{\textrm {T}})}/{\partial \tau_{k,L_k}},$ $\partial({\boldsymbol\xi}_{k,1}$ $\otimes{{\boldsymbol\tau}}_{k,1}^{\textrm {T}})/{\partial \frac{2v_{k,1}}{c}},\cdots,{\partial({\boldsymbol\xi}_{k,L_k}\otimes{{\boldsymbol\tau}}_{k,L_k}^{\textrm{T}})}/{\partial \frac{2v_{k,L_k}}{c}}]$. {It is important to note that the CRB in \cite{stoica1990performance} is derived for 1-D horizontal AOA estimation. We extend this expression for joint range-velocity estimation, which is equivalent to 2-D joint horizontal and vertical AOA estimation. Due to space limitations, we will not provide a detailed mathematical explanation here.}

So far, although we have obtained the CRB expression for ${\bf p}_k$, it is complicated to reveal the quantitative relationship between the subcarrier distribution and the range estimation CRB, as well as the sensing symbol distribution and the velocity estimation CRB, due to the coupling between the range and velocity estimation CRB of different targets. Therefore, for simplicity, we simplify the range and velocity estimation CRB for single-target scenarios. According to \cite{10288116}, for large $N_k$ and $G_k$, ${\textrm{CRB}}^{-1}\!({\bf p}_k)$ is approximately a diagonal matrix. 
Therefore, the CRB derived under single-target scenarios could be utilized in multi-target scenarios. The CRB for the single-target in the $1$st path satisfies
	\begin{sequation}\label{CRB}
		\begin{aligned}
			{\textrm{CRB}}^{-1}\!({\bf p}_k)\!\!=\!\frac{2|\beta_{m,k,1}|^2}{(\sigma_{m,k})^2}  \operatorname{Re}\!\left\{{\bf V}^*\!\left[{\bf I}\!-\!{\bf A}\!\left({\bf A}^*\!{\bf A}\right)^{-1}\!\! {\bf A}^*\right]\!{\bf V}\right\}.
		\end{aligned}
\end{sequation}%
Moreover, ${\bf V}^*\!\left[{\bf I}\!-\!{\bf A}\left({\bf A}^* {\bf A}\right)^{-1}\!\! {\bf A}^*\right] {\bf V}$ can be derived as
\begin{sequation}\label{equ22}
	{\bf V}^*\Big[{\bf I}-\!{\bf A}\left({\bf A}^* {\bf A}\right)^{-1} \!{\bf A}^*\Big] {\bf V}={\bf V}^*{\bf V}-{\bf V}^*{\bf A}\left({\bf A}^* {\bf A}\right)^{-1}\! {\bf A}^*{\bf V}.
\end{sequation}Based on the definitions of ${\bf A}$ and ${\bf V}$, ${\bf A}^*{\bf A}$, ${\bf V}^*{\bf V}$, and ${\bf A}^*{\bf V}$ can be derived as (\ref{equAA}), (\ref{equ24}) and (\ref{equ25})
\begin{equation}\label{equAA}
	{\bf A}^*{\bf A}=G_kN_k,
	\vspace*{-8pt}
\end{equation}
\begin{figure*}[t]
	\begin{equation}\begin{small}\label{equ24}
				{\bf V}^*{\bf V}=\left[
				\begin{matrix}
					\sum_{n_k=1}^{N_k}4\pi^2\Delta f^2G_k\boldsymbol{\zeta}_k[n_k]^2 & 4\pi^2\Delta ff_\textrm{c}T_\textrm{s}\sum_{n_k=1}^{N_k}\boldsymbol{\zeta}_k[n_k]\sum_{g=1}^{G_k}{\boldsymbol{\psi}_k[g]}\\
					4\pi^2\Delta ff_\textrm{c}T_\textrm{s}\sum_{n_k=1}^{N_k}\boldsymbol{\zeta}_k[n_k]\sum_{g=1}^{G_k}{\boldsymbol{\psi}_k[g]} & \sum_{g=1}^{G_k}4\pi^2f_\textrm{c}^2T_\textrm{s}^2N_k{\boldsymbol{\psi}_k[g]}^2
				\end{matrix}\right],
			\end{small}
			\vspace*{-12pt}
	\end{equation}
\end{figure*}
\begin{equation}\label{equ25}
		\begin{small}
			{\bf V}^*{\bf A}=\left[
			\begin{matrix}
				j2\pi\Delta fG_k\sum\limits_{n_k=1}^{N_k}\boldsymbol{\zeta}_k[n_k],	j2\pi f_cT_\textrm{s}N_k\sum\limits_{g=1}^{G_k}\boldsymbol{\psi}_k[g]
			\end{matrix}\right]^{\textrm{T}}\!\!.	
		\end{small}
\end{equation} Then, by substituting (\ref{equAA}) and (\ref{equ25}) into (\ref{equ22}), ${\bf V}^*{\bf A}\left({\bf A}^*\! {\bf A}\right)^{-1}\!\! {\bf A}^*{\bf V\!}$ can be obtained as (\ref{equ26}).
\begin{figure*}
	\begin{equation}\label{equ26}
			\setlength{\arraycolsep}{0.8pt}
			\begin{small}
				{\bf V}^*{\bf A}\left({\bf A}^* {\bf A}\right)^{-1} {\bf A}^*{\bf V}=\frac{1}{G_kN_k}\!\left[
				\begin{matrix}
					4\pi^2\Delta f^2G_k^2[\sum_{n_k=1}^{N_k}\boldsymbol{\zeta}_k[n_k]]^2 & 4\pi^2\Delta ff_\textrm{c}T_\textrm{s}G_kN_k\sum_{n_k=1}^{N_k}\boldsymbol{\zeta}_k[n_k]\sum_{g=1}^{G_k}{\boldsymbol{\psi}_k[g]}\\
					4\pi^2\Delta ff_\textrm{c}T_\textrm{s}G_kN_k\sum_{n_k=1}^{N_k}\boldsymbol{\zeta}_k[n_k]\sum_{g=1}^{G_k}{\boldsymbol{\psi}_k[g]} & 4\pi^2f_\textrm{c}^2T_\textrm{s}^2N_k^2[\sum_{g=1}^{G_k}{\boldsymbol{\psi}_k[g]}]^2
				\end{matrix}\right]\!.
			\end{small}
			\vspace*{-8pt}
\end{equation}
\end{figure*} 
Consequently, ${\bf V}^*\!\!\left[{\bf I}\!-\!{\bf A}\left({\bf A}^* {\bf A}\right)^{-1}\!\! {\bf A}^*\right]\! {\bf V}$ can be formulated as
\begin{sequation}\label{equ27}
		\setlength{\arraycolsep}{1.6pt}
		\begin{aligned}
			&{\bf V}^*\!\!\left[{\bf I}\!-\!\!{\bf A}\left({\bf A}^*\!{\bf A}\right)^{-1}\!{\bf A}^*\right] \!\!{\bf V}\! =4\pi^2G_kN_k\!\!\left[
			\begin{matrix}
				\Delta f^2\bar{\boldsymbol\zeta}_k\! &\! 0\\
				0 \!&\!  f_\textrm{c}^2T_\textrm{s}^2\bar{\boldsymbol{\psi}}_k
			\end{matrix}\right]\!,
		\end{aligned}
\end{sequation}%
where $\bar{\boldsymbol\zeta}_k$ and $\bar{\boldsymbol{\psi}}_k$ represent the variance of the set $\{{\boldsymbol\zeta}_k[1],\cdots,{\boldsymbol\zeta}_k[N_k]\}$ and $\{{\boldsymbol{\psi}_k[1]},\cdots,{\boldsymbol{\psi}_k[G_k]}\}$, respectively. Moreover, $\bar{\boldsymbol\zeta}_k$ and $\bar{\boldsymbol{\psi}}_k$ are respectively expressed as
\begin{subequations}
	\begin{small}
		\label{equ291}
		\begin{align}
			&\bar{\boldsymbol\zeta}_k={1}/{N_k}\sum\nolimits_{n_k=1}^{N_k}\!\boldsymbol{\zeta}_k[n_k]^2\!-{1}/{N_k^2}\left[\sum\nolimits_{n_k=1}^{N_k}\!\boldsymbol{\zeta}_k[n_k]\right]^2,\\
			&\bar{\boldsymbol{\psi}}_k={1}/{G_k}\sum\nolimits_{g=1}^{G_k}{\boldsymbol{\psi}}_k[g]^2-{1}/{G_k^2}\left[\sum\nolimits_{g=1}^{G_k}{\boldsymbol{\psi}}_k[g]\right]^2.
		\end{align}
	\end{small}
\end{subequations}

In the derivation process, it is important to note that, if we exclude $\beta_{m,k,1}$ when deriving the CRB for range and velocity, $\bar{\boldsymbol\zeta}_k$ and $\bar{\boldsymbol{\psi}}_k$ become ${1}/{N_k}\sum_{n_k=1}^{N_k}\boldsymbol{\zeta}_k[n_k]^2$ and ${1}/{G_k}\sum_{g=1}^{G_k}{\boldsymbol{\psi}_k[g]}^2$, respectively. \textit{In this case, the values of ${1}/{N_k}\sum_{n_k=1}^{N_k}\boldsymbol{\zeta}_k[n_k]^2$ and ${1}/{G_k}\sum_{g=1}^{G_k}{\boldsymbol{\psi}_k[g]}^2$ will vary with the change of the logical reference points in steering vectors ${\boldsymbol\tau}_{k,1}$ and ${\boldsymbol\xi}_{k,1}$, even if the physical antenna structure remains constant}. $\bar{\boldsymbol\zeta}_k$ and $\bar{\boldsymbol{\psi}}_k$ will be numerically equivalent to those in (\ref{equ291}) only if the reference points are chosen at the geometric centres of the vectors ${\boldsymbol\tau}_{k,1}$ and ${\boldsymbol\xi}_{k,1}$. This phenomenon is obviously inconsistent with the actual situation. 
Specifically, we provide the explanations for this phenomenon.

\textbf{Remark:} In actual sensing systems, for the $k$th UE, the CRB for range and velocity are determined by all of $\beta_{m,k,1}$, ${\boldsymbol\tau}_{k,1}$, and ${\boldsymbol\xi}_{k,1}$. Here we take the range estimation CRB as an example. When a new logical reference point is chosen in ${\boldsymbol\tau}_{k,1}$, both $\beta_{m,k,1}$ and the ${\boldsymbol\tau}_{k,1}$ undergo additional and opposite phase rotations, for example  $e^{jp_{\textrm{e}}}{\boldsymbol\tau}_{k,1}$ and $e^{-jp_{\textrm {e}}}\beta_{m,k,1}$. 
Then, if $\beta_{m,k,1}$ is included in the CRB derivations, since $e^{jp_{\textrm{e}}}e^{-jp_{\textrm{e}}}=1$, the rotation does not impact the signal model in (\ref{compensation1}) at all. Consequently, the CRB remains unaffected, and (\ref{equ291}) will always hold. However, if $\beta_{m,k,1}$ is not included, the phase rotation $e^{jp_{\textrm{e}}}$ on $\beta_{m,k,1}$ will be disregarded, while the rotation on ${\boldsymbol\tau}_{k,1}$ exists. This leads to an additional phase rotation in the signal model (\ref{compensation1}). Therefore, the CRB for range and velocity changes with alterations of the reference point, which is unrealistic.

Then, according to (\ref{CRB}), (\ref{equ27}), (\ref{equ291}), as well as the relationship $R = c\tau_{k,1}$ and $v =\frac{2v_{k,1}}{c}\times \frac{c}{2}$, the CRB for range and velocity can be further formulated as 
\begin{equation}\label{equ29}
		{\textrm {CRB}}(R)={c^2(\sigma_{m,k})^2}/({8|\beta_{m,k,1}|^2\pi^2G_kN_k\Delta f^2\bar{\boldsymbol\zeta}_k}),
	\end{equation}
	\begin{equation}\label{equ29_1}
		{\textrm {CRB}}(v)={c^2(\sigma_{m,k})^2}/({32|\beta_{m,k,1}|^2\pi^2G_kN_kf_\textrm{c}^2T_\textrm{s}^2\bar{\boldsymbol{\psi}}_k}).
\end{equation}
Then we can get the insight that \textit{the CRB for range and velocity are inversely proportional to the variance of the subcarrier indexes and sensing symbol indexes, respectively.} Moreover, since the range estimation CRB and the estimation velocity CRB are decoupled, they can be individually optimized by configuring the utilized subcarriers and sensing symbols.

\subsection{CRB Optimization in Single-UE OFDMA PMNs}
In this subsection, we optimize ${\textrm{CRB}}(R)$ with respect to the variance of subcarrier indexes and ${\textrm{CRB}}(v)$ with respect to the variance of sensing symbol indexes, respectively, in single-UE OFDMA PMNs.

Specifically, the integer optimization problems for ${\textrm{CRB}}(R)$ and ${\textrm{CRB}}(v)$ are formulated as (\ref{equ23}) and (\ref{equ24_1}), respectively. 
According to (\ref{equ23}) and (\ref{equ24_1}), the optimization problem of ${\textrm{CRB}}(R)$ with regard to the variance of the subcarrier indexes is the same as that of ${\textrm{CRB}}(v)$ with regard to the variance of the sensing symbol indexes. Therefore, we only provide the proof for the ${\textrm{CRB}}(R)$ optimization. It is important to note that the conclusions obtained are equally applicable to the optimization of ${\textrm{CRB}}(v)$. 
\begin{equation}\label{equ23}
		\begin{aligned}
			\mathop{\min}_{{\boldsymbol{\zeta}}_k} & \ \  {c^2(\sigma_{m,k})^2}/({8|\beta_{m,k,1}|^2\pi^2G_kN_k\Delta f^2\bar{\boldsymbol\zeta}_k})\\
			{\textrm{s.t.}}&\ \  
			{\operatorname{card}}\{{\boldsymbol{\zeta}}_k[n]|n=1,\cdots,\!N_k\}=\!N_k,\\
			&\ \ {\boldsymbol{\zeta}}_k\in\!\{1,\!\cdots,\!N\}^{1\times N_k}, N_k\le N,\ N_k\in\mathbb{Z}.
		\end{aligned}
	\end{equation}
	\begin{equation}\label{equ24_1}
		\begin{aligned}
			\mathop{\min}_{{\boldsymbol{\psi}}_k} & \ \  {c^2(\sigma_{m,k})^2}/({32|\beta_{m,k,1}|^2\pi^2G_kN_k f_\textrm{c}^2T_\textrm{s}^2\bar{\boldsymbol{\psi}}_k})\\
			{\textrm {s.t.}}
			& \ \  			{\operatorname{card}}\{{\boldsymbol{\psi}}_k[g]|\ g=1,\cdots,\!G_k\}\!=\!G_k,\\
			& \ \ {\boldsymbol{\psi}}_k\!\in\!\{1,\!\cdots,\!G\}^{1\times G_k},G_k\le G,\ G_k\in\mathbb{Z}.
		\end{aligned}
\end{equation}

In particular, the optimal distribution is depicted as Proposition \ref{pro1_1}.

\begin{proposition}\label{pro1_1}
	$\frac{1}{N_k}\sum_{n_k=1}^{N_k}\boldsymbol{\zeta}_k[n_k]^2-\frac{1}{N_k^2}[\sum_{n_k=1}^{N_k}\boldsymbol{\zeta}_k[n_k]]^2$ is maximized and \rm{${\textrm{CRB}}(R)$} \textit{is minimized, when subcarriers are assigned as the edge-first distribution, where subcarriers near the edge are preferentially assigned to the UE.}
\end{proposition}

\textit{Proof}: See Appendix \ref{Pro3}.

Intuitively, the variance of a set of numbers reflects the amount of dispersion among the numbers. Therefore, maximizing ${\textrm{CRB}}(R)$ implies that the subcarrier indexes should be distributed as 'dispersed' as possible. Proposition \ref{pro1_1} provides a criterion for achieving maximum dispersion: prioritize indexes closer to the edges, assuming the elements are arranged in ascending or descending order.

Moreover, we also determine the worst subcarrier distribution in Proposition \ref{pro1}.
\begin{figure}[t]
	\centering
	\includegraphics[width=3.5in]{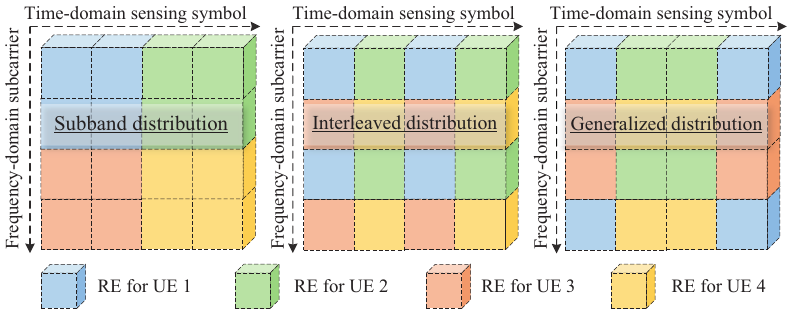}
	\caption{Several common subcarrier and sensing symbol distribution schemes. Here, 'RE' stands for 'resource element'. Due to space limitations, the edge-first distribution is illustrated as the distribution scheme for UE 1 under the ``generalized distribution".}
	\label{CAS}
	\vspace*{-6pt}
\end{figure}
\begin{proposition}\label{pro1}
$\frac{1}{N_k}\sum_{n_k=1}^{N_k}\boldsymbol{\zeta}_k[n_k]^2-\frac{1}{N_k^2}[\sum_{n_k=1}^{N_k}\boldsymbol{\zeta}_k[n_k]]^2$ is minimized and {\rm${\textrm{CRB}}(R)$} is maximized, when subcarriers are assigned as the subband distribution. 
\end{proposition}

\textit{Proof}: See Appendix \ref{Pro2}.

Proposition \ref{pro1} implies that to minimize the variance of the subcarrier indexes and thereby maximize the CRB, one should distribute the subcarriers as closely as possible, assuming that the subcarrier indexes are arranged either from small to large or from large to small.

\textbf{Remark:} Since the mathematical relationship between range estimation CRB and $\boldsymbol{\zeta}_k$ is identical to that between velocity estimation CRB and $\boldsymbol{\psi}_k$, the optimization for sensing symbols  also adheres to the criterion outlined in Propositions 2 and 3.

So far, we have identified the best and worst subcarrier and sensing symbol distribution for range and velocity estimation in the single-UE OFDMA PMNs. We then optimize distribution in more complex multi-UE OFDMA PMNs.

\subsection{CRB Optimization in Multi-UE OFDMA PMNs}
In this subsection, we mainly focus on optimizing subcarrier distribution for improving range estimation CRB for an entire OFDMA PMN\footnote{The optimal sensing symbol distribution in multi-UE OFDMA PMNs is similar to that in single-UE OFDMA PMNs because, in multi-UE scenarios, sensing symbols belonging to different UEs can overlap in the same time resources.}. To ensure that each UE has acceptable performance, we aim to maximize the minimum variance of the integer subcarrier index groups of $K$ UEs, i.e. to minimize the maximum CRB of all UEs. Specifically, the optimization problem is formulated as
\begin{sequation}\label{equ30}
	\begin{aligned} 
		\mathop{\max}\limits_{{\bf \Xi}} & \  \mathop{\min}\limits_{k}\  {\frac{1}{N_k}{\boldsymbol{\zeta}}_k{\bf \Gamma}_k{\bf S}({\boldsymbol{\zeta}}_k{\bf \Gamma}_k)^{\textrm {T}}-\frac{1}{N^2_k}||{\boldsymbol{\zeta}}_k{\bf \Gamma}_k{\boldsymbol \chi}||_2}, \\
		{\textrm{s.t.}}& \ \ k\in\{1,\cdots,K\}, {\bf \Xi}{\bf \Xi}^{\textrm {T}}=N_k{\bf I}_N, {\bf \Xi}\in\{0,1\}^{N\times K},
	\end{aligned}
	\end{sequation}%
	where ${\boldsymbol \chi}=[1,\cdots,N]^{\textrm {T}}$, ${\bf S}={\textrm {diag}}(1^2,\cdots,N^2)$ and ${\bf \Xi}=[{\bf \Gamma}_1^{\textrm {T}}{\boldsymbol{\zeta}}_1^{\textrm {T}},\cdots,{\bf \Gamma}_K^{\textrm {T}}{\boldsymbol{\zeta}}_K^{\textrm {T}}]$.
	
	The problem in (\ref{equ30}) is essentially a kind of cardinality-constrained MWNP problem \cite{JournalofScheduling}. In particular, it involves assigning a certain number of non-overlapping subcarriers to $K$ UEs, i.e. partitioning the subcarrier index set $\{1, \cdots, N\}$ into $K$ subsets without any shared elements. The goal of this partitioning is to maximize the minimum variance of $K$ subsets. 
	This is an NP-hard problem \cite{JournalofScheduling}, and finding the optimal solution results in extremely high computational complexity. Moreover, in practical mobile scenarios, both $N$ and $K$ are often large (e.g., $N=1024$ and $K=20$), 
	making it difficult to obtain the globally optimal solution directly. 
	
	As a result, we aim to find a near-optimal solution by leveraging inequalities, which can significantly reduce computational complexity. Specifically, in Proposition \ref{pro4}, we present the upper bound for the problem and the conditions that distribution schemes should satisfy to achieve this bound.
	\begin{proposition}\label{pro4}
	Assume that the maximized minimum variance $\epsilon$ is achieved by the {\rm$k_{\textrm{m}}$th} index subset. Then, $\epsilon$ satisfies
	\begin{equation}\label{equ31}
		\epsilon\le{1}/{(KN_{k_{\textrm {m}}})}\sum\nolimits_{n=1}^{N}(n-\bar{n})^2=N\epsilon_\textrm{t}/(KN_{k_\textrm{m}}),
	\end{equation}%
	where $\bar{n}={1}/{N}\sum_{n=1}^{N}n$ and $\epsilon_\textrm{t}=1/N\sum\nolimits_{n=1}^{N}(n-\bar{n})^2$ is the variance of the entire set. The equality holds if and only if the mean and variance of the $K$ partitioned subsets are identical.
	\end{proposition}

	\textit{Proof}: See Appendix \ref{proof4}.

	Essentially, the proposition articulates an intuitive and fundamental principle: the variance of any subcarrier index subset cannot exceed the bound affected by the variance of the entire subcarrier index set.
	Furthermore, the upper bound of the minimized maximum CRB for range estimation, denoted as $\textrm{CRB}_\textrm{low}(R)$, can be expressed as 
	\begin{equation}\label{equ34_1}
	\textrm{CRB}_\textrm{low}(R)=\frac{Kc^2(\sigma_{m,k})^2}{8|\beta_{m,k,1}|^2\pi^2G_k\Delta f^2\sum_{n=1}^{N}(n-\bar{n})^2}.
	\end{equation}
	However, finding a particular partition scheme that ensures the same mean and variance among $K$ subcarrier index subsets, to achieve the lower bound is still difficult, due to the extremely high complexity.

	To address this problem, we aim to find a solution to (\ref{equ30}) with an additional constraint concerning fair resource allocation among multiple UEs\footnote{This additional constraint is quite special because the CRB of the optimal solution to (\ref{equ30}) with this constraint is very close to the CRB of the globally optimal solution to (\ref{equ30}) without the constraint, as proved in Proposition \ref{pro6}.}. This constraint is designed to satisfy the condition in Proposition \ref{pro4} as closely as possible. 
	Specifically, we formulate the constraint as
	\begin{equation}\label{equ34}
	{\boldsymbol\zeta}_k[n]={\boldsymbol\zeta}_{k^{'}}[n]+\eta_{k,k^{'}},
	\end{equation} where $n\in\{1,\cdots,N_k\}$, $k,k^{'}\in\{1,\cdots,K\}$ and $k\neq k^{'}$. Moreover, $\eta_{k,k^{'}}$ is an constant integer and $\eta_{k,k^{'}}\neq 0$.
	Since this constraint essentially ensures the same variance among the $K$ subsets according to (\ref{equ34}), it partially meets the condition in Proposition \ref{pro4} and ensures that all UEs achieve the same sensing performance. Then, Proposition \ref{pro5} is presented to identify the optimal partition scheme under the additional constraint.
	\begin{proposition}\label{pro5}
	under the constraint
	\begin{equation}\label{equ32}
		{\boldsymbol\zeta}_k[n]={\boldsymbol\zeta}_{k^{'}}[n]+\eta_{k,k^{'}},
	\end{equation} the minimum variance of the $K$ index subsets is maximized, when $\eta_{k,k^{'}}=k^{'}-k$.
	\end{proposition}
	\textit{Proof}: See Appendix \ref{proof5}.
	\begin{figure}[t]
	\centering
	\includegraphics[width=2.9in]{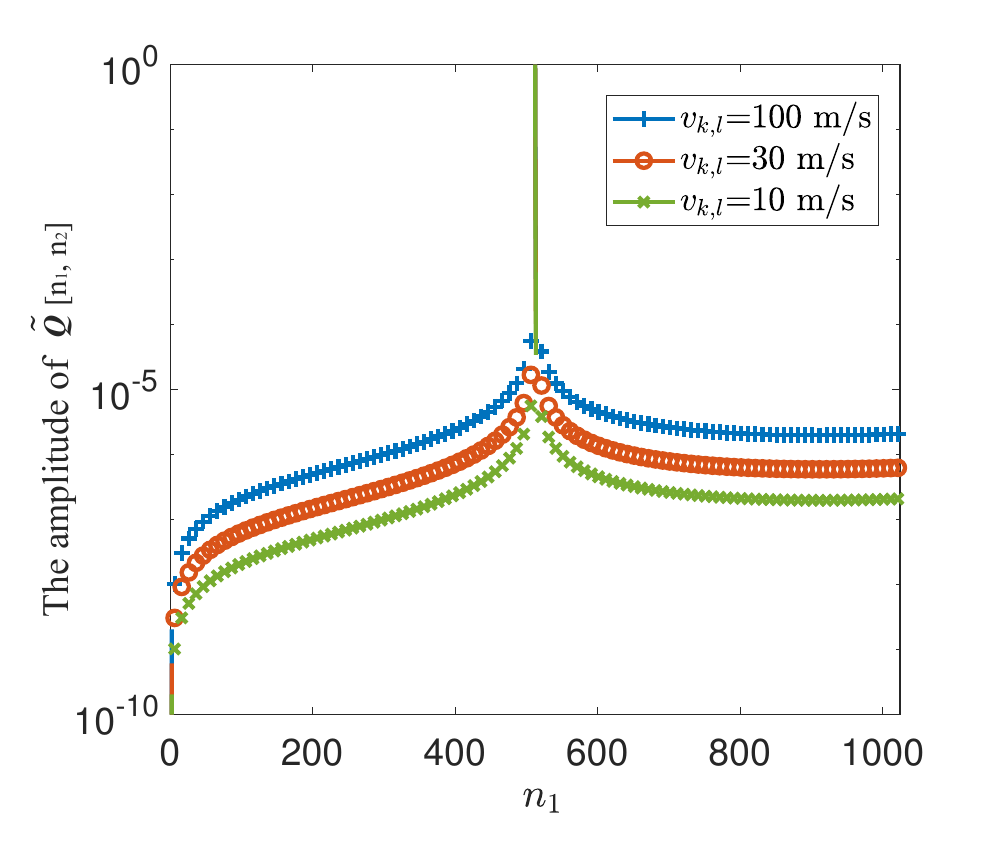}
	\caption{The ICI caused by the $n_1$th subcarrier on the $n_2$th subcarrier, where $n_1 \neq n_2$, under different velocities. For simplicity, $v_{k,l}$ for $k=1,\cdots,K$ and $l=1,\cdots,L_k$ are set as $v$ in the legend. }\label{figure4}
	\vspace*{-6pt}
	\end{figure}

	Proposition \ref{pro5} implies that optimal sensing performance under the constraint is achieved when the variance of the means of the $K$ subsets is minimized. At this point, $\eta_{k,k^{'}}=k^{'}\!-k$ for the $k$th and $k^{'}$th UEs, indicating that the \textit{interleaved subcarrier distribution} results in optimal sensing performance. 

	Furthermore, to evaluate the performance of the CRB for the interleaved distribution, in Proposition \ref{pro6}, we provide a theoretical gap between $\textrm{CRB}_\textrm{low}(R)$ and the CRB for the interleaved distribution, which is denoted as $\textrm{CRB}_\textrm{I}(R)$. The gap is quantified using the metric $[{\textrm{CRB}_\textrm{I}(R)-\textrm{CRB}_\textrm{low}(R)}]/{\textrm{CRB}_\textrm{low}(R)}$.
	\begin{proposition}\label{pro6}
	{\rm$[{\textrm{CRB}_\textrm{I}(R)-\textrm{CRB}_\textrm{low}(R)}]/{\textrm{CRB}_\textrm{low}(R)}$} is derived as ${(K\!\!-\!\!1)(K\!\!+\!\!1)}/\left[{(N\!\!-\!\!1)(N\!+\!1)}\!-\!{(K\!\!-\!\!1)(K\!\!+\!\!1)}\right]$.
	\end{proposition}

	\textit{Proof}: See Appendix \ref{proof6}.

	Since $N$ and $K$ are generally large, $\frac{\textrm{CRB}_\textrm{I}(R)-\textrm{CRB}_\textrm{low}(R)}{\textrm{CRB}_\textrm{low}(R)}$ is generally very small according to Proposition \ref{pro6}. For instance, in a scenario with $N=1024$ and $K=20$, $\frac{\textrm{CRB}_\textrm{I}(R)-\textrm{CRB}_\textrm{low}(R)}{\textrm{CRB}_\textrm{low}(R)}$ is approximately $0.000354$, which can be neglected. Thus, the CRB corresponding to the interleaved distribution almost approaches the CRB lower bound in practical deployments.

\section{The Achievable Communication Rates Under Different Distribution Schemes}
	In this section, we explore how various subcarrier distribution schemes impact communication performance by formulating the theoretical achievable communication rates in narrowband synchronous systems. Our formulation reveals that subcarrier distribution schemes affect the achievable communication rates by influencing the power of ICI. However, a theoretical examination shows that the power of ICI under general distribution schemes is negligible, thereby having a minimal impact on the achievable communication rates. 

	First, by reviewing (\ref{y_g^n(t)}), we reformulate the $p$th sample of the $g$th OFDM symbol received by $M_{\textrm{R}}$ antennas, namely ${\bf y}_{g}(pT_{\textrm{sam}})\in\mathbb{C}^{M_{\textrm{R}}\times 1}$, as follows
	\begin{sequation}\label{y_g^n(t)1}
	\begin{aligned}
		{\bf y}_{g}(pT_{\textrm{sam}})\!=\!\sum\limits_{k=1}^{K}\sum\limits_{l=1}^{L_k}\alpha_{k,l}{\bf a}(\Omega_{k,l}^{\textrm {r}}){\bf a}^{\textrm {T}}(\Omega_{k,l}^{\textrm{t}}){\boldsymbol \varpi}_k{\bf s}_{g,k}{\bf q}_{k,l,p}\!+\!{\bf w}_g(pT_{\textrm{sam}}\!),
	\end{aligned}
	\end{sequation}%
	where ${\bf q}_{k,l,p}[n]=e^{-j2\pi (n-1)\Delta f[(\frac{2v_{k,l}}{c}-1)pT_{\textrm{sam}}+\tau_{k,l}]}$ for $n=1,\cdots, N$. The $g$th OFDM symbol received by $M_{\textrm{R}}$ antennas, ${\bf Y}_{g}=[{\bf y}_{g}(T_{\textrm{sam}}),\cdots,$ ${\bf y}_{g}(NT_{\textrm{sam}})]$, can be reformulated as
	\begin{sequation}
	{\bf Y}_{g}\!=\!\sum\nolimits_{k=1}^{K}\sum\nolimits_{l=1}^{L_k}\!\alpha_{k,l}{\bf a}(\Omega_{k,l}^{\textrm{r}}){\bf a}^{\textrm {T}}\!(\Omega_{k,l}^{\textrm{ t}}){\boldsymbol \varpi}_k{\bf s}_{g,k}{\bf Q}_{k,l}\!+\!{\bf W}_g,
	\end{sequation}%
	where ${\bf Q}_{k,l}=[{\bf q}_{k,l,1},\cdots,{\bf q}_{k,l,N}]$. We then transform ${\bf Y}_{g}$ into frequency domain, and the signals on $N_{k^{'}}$ subcarriers of the $k^{'}$th UE can be expressed as
	\begin{fequation}\label{Y_g_1}
		\tilde{\bf Y}_{g,k^{'}}\!=\!{\bf Y}_{g}{\bf F}_N{\boldsymbol{\Gamma}_{k^{'}}^{\textrm {T}}}\!=\!\sum\limits_{k=1}^{K}\!\!\Big[\Big(\sum\limits_{l=1}^{L_k}{\bf h}_{k,l}{\bf s}_{g,k}{\bf Q}_{k,l}\Big){\bf F}_N\Big]{\boldsymbol{\Gamma}_{k^{'}}^{\textrm{T} }}\!+\!\tilde{\bf W}_{g,{k^{'}}},
	\end{fequation}%
	where ${\bf h}_{k,l}$ is defined as $\alpha_{k,l}{\bf a}(\Omega_{k,l}^{\textrm{r}}){\bf a}^{\textrm{T}}(\Omega_{k,l}^{\textrm {t}}){\boldsymbol \varpi}_k$ and $\tilde{\bf W}_{g,{k^{'}}}={\bf W}_{g,{k^{'}}}{\bf F}_N{\boldsymbol{\Gamma}_{k^{'}}^{\textrm{T}}}$. 

	To evaluate the achievable communication rates, we first split $\tilde{\bf Y}_{g,{k^{'}}}$ as\begin{fequation}\label{equ381}
	\begin{aligned}
			&\tilde{\bf Y}_{g,{k^{'}}\!}=\\
			&\underbrace{\!\sum\limits_{l=1}^{L_{k^{'}}}\!{\bf h}_{k^{'}\!,l}{\bf s}_{g,k^{'}}\!{\boldsymbol{\Gamma}_{k^{'}}^{\textrm {T}}}}_{\textrm{desired signal}} 
			\!\!+\!\underbrace{\Big[\!\!\sum\limits_{\substack{k\neq k^{'}\!\!,  k=1 }}^{K}\!\!\sum\limits_{l=1}^{L_k}\!{\bf h}_{k\!,l}{\bf s}_{g\!,k}{\bf Q}_{k\!,l}{\bf F\!}_N\!-\!\!\sum\limits_{l=1}^{L_{k^{'}}}\!{\bf h}_{k^{'}\!\!,l}{\bf s}_{g\!,k^{'}}\!\!\Big]{\boldsymbol{\Gamma}_{k^{'}}^{\textrm {T}}}}_{\textrm{ICI}}
			\!+\! \underbrace{\tilde{\bf W}_{g\!,{k^{'}\!}}}_{\textrm{noise}}\!.
	\end{aligned}
	\end{fequation}%
	Consequently, the achievable communication rates of the $n$th subcarrier (assigned to the $k^{'}$th UE), denoted as $R_{g,{k^{'}},n}$, can be formulated as
	\begin{sequation}
	R_{g,{k^{'}}\!\!,n}\!\!=\! \operatorname{log}\!\Big(\!\det( {\bf I}_{M_{\textrm {R}}}\!\!+\!\!\frac{(\sum\nolimits_{l=1}^{L_{k^{'}}}\!{\bf h}_{k^{'}\!\!,l}{\bf s}_{g,k^{'}}\![n])(\sum\nolimits_{l=1}^{L_{k^{'}}}\!{\bf h}_{k^{'}\!\!,l}{\bf s}_{g,k^{'}}\![n])^{\textrm{H}}}{\breve{\sigma}^2_{g,n}})\Big),
	\end{sequation}%
	where $\breve{\sigma}^2_{g,n}$ represents the interference-plus-noise power on the $n$th subcarrier.
	
	Furthermore, considering that the power of ICI and noise is statistically identical over $M_{\textrm{R}}$ antennas, and the noise is statistically independent to ICI, for signals received by the $m$th antenna, $m=1,\cdots,M_\textrm{R}$, $\breve{\sigma}^2_{g,n}$ can be formulated as
	\begin{fequation}
				\breve{\sigma}^2_{g,n}
				\!=\!{\mathrm E}\Big(\Big\Vert\Big[\sum_{k=1}^{K}\!\sum_{l=1}^{L_{k}}\!{\bf h}_{k,l}[m]\tilde{\bf s}_{g,k,l}[n]-\!\sum_{l=1}^{L_{k^{'}}}{\bf h}_{k^{'},l}[m]{\bf s}_{g,k^{'}}[n]\Big\Vert^2\Big)+\tilde\sigma_{g,{k^{'}}}^2,
	\end{fequation}%
	where $\tilde{\bf s}_{g,k,l}={\bf s}_{g,k}{\bf Q}_{k,l}{\bf F}_N$ and $\tilde\sigma_{g,{k^{'}}}^2$ is the noise power.
	
	Specifically, based on the expression of ${\bf q}_{k,l,p}$ and ${\bf F}_N$, the $(n_1,n_2)$th element of the ICI matrix $\tilde{\bf Q}_{k,l}={\bf Q}_{k,l}{\bf F}_N$ can be formulated as 
	\begin{fequation}\label{equ43}
	\begin{aligned}
			&\tilde{\bf Q}_{k,l}[n_1,n_2]\\
			&=\!{\sum\nolimits_{p=1}^{N}\!e^{-j2\pi \{(n_1\!-\!1)\Delta f[({2v_{k,l}}/{c}-1)(p-1)T_{\textrm {sam}}+ \tau_{k,l}]+ \frac{(n_2-1)(p-1)}{N}\}}}\!/\!{N}\\
			&=\!e^{-\!j2\pi (n_1\!-\!1)\Delta f\tau_{k,l}}/N\sum\nolimits_{p=1}^{N}e^{j2\pi \frac{[(1-{2v_{k,l}/{c})n_1-n_2+2v_{k,l}}/{c}](p-1)}{N}}\\
			&=\!e^{-\!j2\pi (n_1\!-\!1)\Delta f\tau_{k,l}}/N\!\sum\nolimits_{p=1}^{N} \!e^{j2\pi \frac{(n_1-n_2)(p-1)}{N}}\!e^{j2\pi \frac{{2v_{k,l}}/{c}(1-n_1)(p-1)}{N}}\!.
	\end{aligned}
	\end{fequation}%
	Since $\frac{2v_{k,l}}{c}$ is sufficiently small, the term $e^{j2\pi \frac{\frac{2v_{k,l}}{c}(1-n_1)(p-1)}{N}}$ in (\ref{equ43}) can be approximated as 1. Therefore, we have $\tilde{\bf Q}_{k,l}\approx {\bf I}_N$, indicating $\breve{\sigma}^2_{g,n}=\tilde\sigma_{g,{k^{'}}}^2$. 
	This suggests that the ICI can be disregarded under different distribution schemes. 
	Therefore, the distribution schemes have almost no effect on the achievable communication rates. Moreover, according to (\ref{equ43}), the sensing symbol index $g$ does not affect $\tilde{\bf Q}_{k,l}[n_1,n_2]$. Thus, the sensing symbol distribution does not affect the achievable communication rates. 
	
	To demonstrate the negligible impact of ICI, we present ICI in Fig. \ref{figure4} using a system with $N=1024$ as an example. 
	It is evident that the ICI is remarkably small, on the order of $10^{-5}$. Compared to the unitary power of the signals, the power of the ICI is negligible.

\begin{figure*}[t]
	\centering
	\subfigure[Subband distribution] {\includegraphics[width=2.25in]{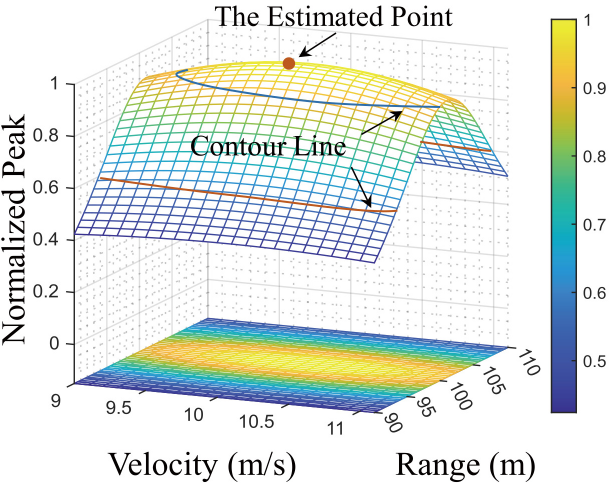}}\label{f1}
	\subfigure[Interleaved distribution] {\includegraphics[width=2.25in]{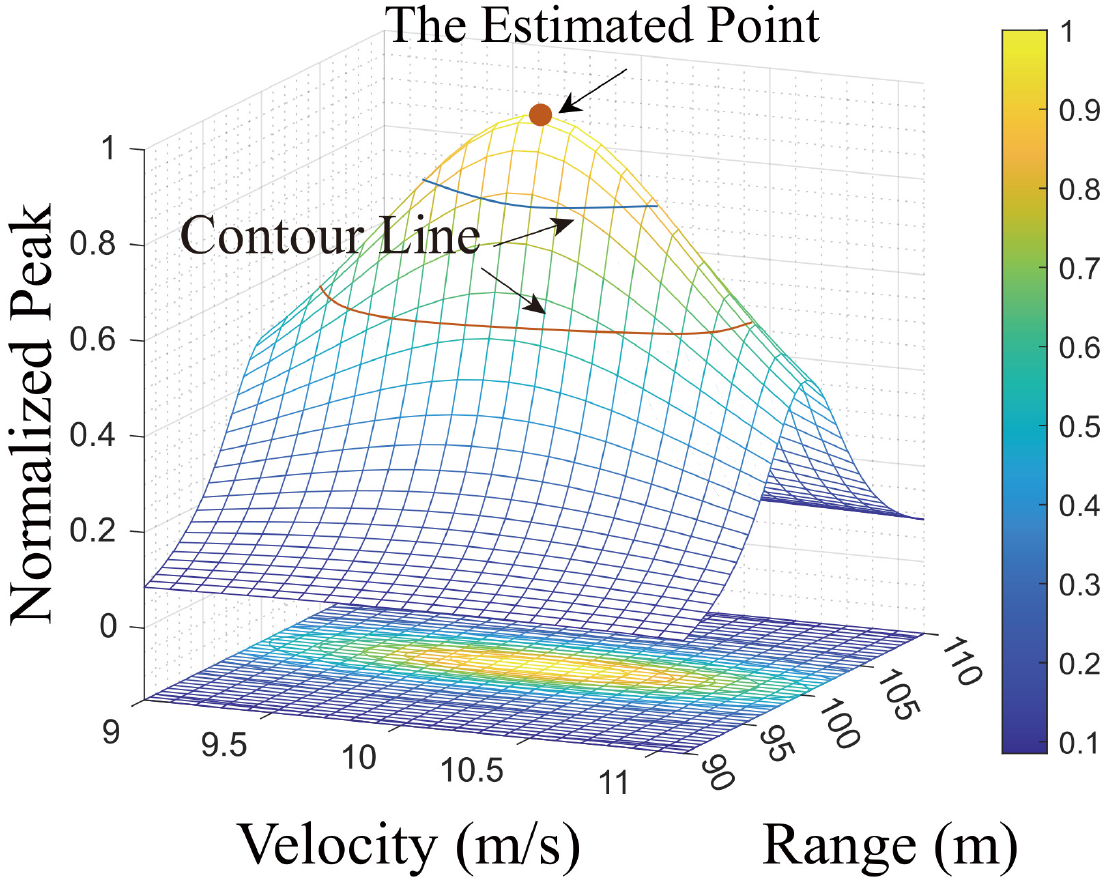}}\label{f2}
	\subfigure[Edge-first distribution] {\includegraphics[width=2.25in]{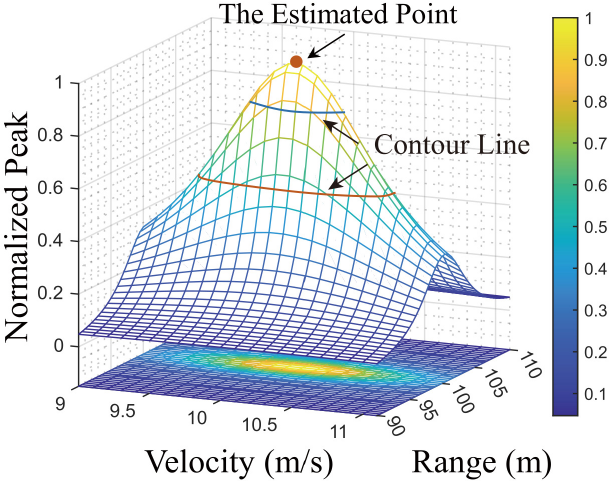}}\label{f3}
	\caption{The range and velocity estimation peaks under different subcarrier and sensing symbol distribution schemes.}
	\label{fig4}
\end{figure*}

\section{Numerical Simulations}

In this section, numerical simulations are performed to evaluate our theoretical analysis. 
Specifically, we set the number of the receiving and transmitting antennas to $M_{\textrm{R}}=8$ and $M_{\textrm{T}}=2$, respectively. Moreover, the carrier frequency $f_\textrm{c}$ is assumed to be $28$ GHz, while the subcarrier spacing is chosen to be $100$ kHz, as assumed in \cite{ni2021uplink,rahman2019framework}. 

To intuitively compare the range and velocity estimation performance under different distribution schemes, we present the estimation peaks in Fig. \ref{fig4}. Specifically, we set up a single UE assigned with 16 subcarriers and 16 sensing symbols in the simulation. The subcarriers and sensing symbols are selected from a pool of 48 consecutive subcarriers and 48 consecutive sensing symbols, respectively. Additionally, we assume a signal-to-noise ratio (SNR) of 20 dB in the simulation. 
As depicted in Fig. \ref{fig4}, the estimation peak under the interleaved distribution is sharper than that under the subband distribution, while the peak under the edge-first distribution is sharper than that under the interleaved distribution. Since a sharper estimation peak corresponds to a better CRB according to \cite{wxy-1}, the edge-first scheme outperforms both the subband and interleaved schemes, and the interleaved scheme outperforms the subband scheme, as described in our theoretical analysis. 

\begin{figure}[htb]
	\centering
	\includegraphics[width=2.9in]{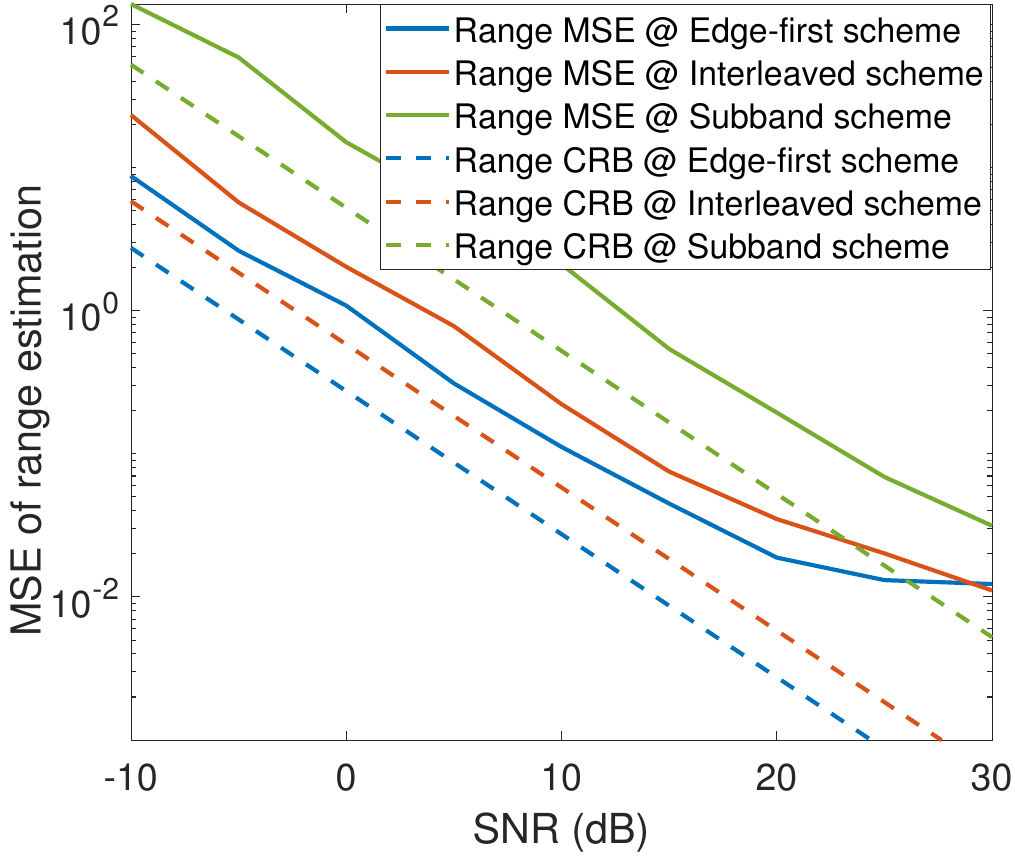}
	\caption{Range estimation MSE and CRB under different subcarrier distribution schemes in single-UE OFDMA PMNs.}\label{figure5}
\end{figure}
\begin{figure}[htb]
	\centering
	\includegraphics[width=2.9in]{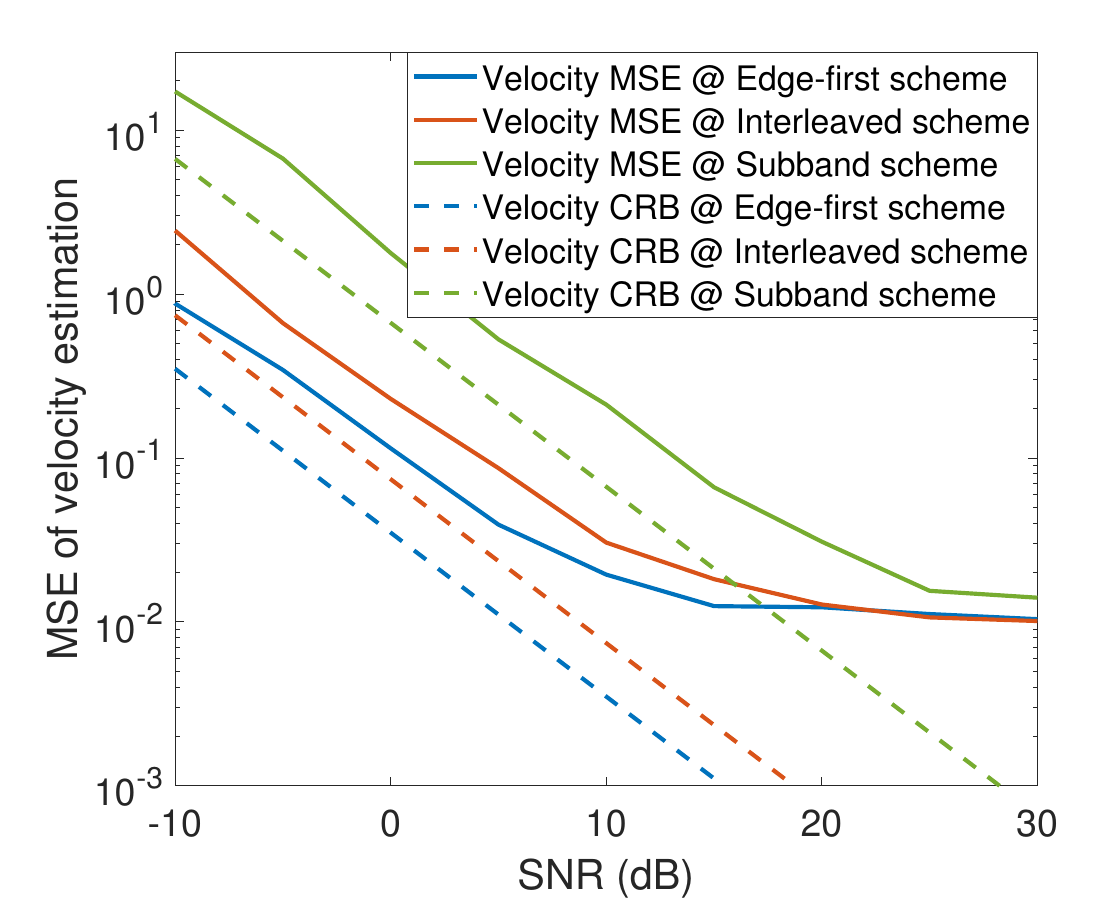}
	\caption{Velocity estimation MSE and CRB under different subcarrier distribution schemes in single-UE OFDMA PMNs.}\label{figure6}
\end{figure}

In single-UE OFDMA PMNs, the mean squared error (MSE) and CRB for range and velocity estimation under different subcarrier and sensing symbol distribution are presented in Fig. \ref{figure5} and Fig. \ref{figure6}, respectively. 
In the simulation, the setting for the single UE is the same as that in the simulation of Fig. \ref{fig4}.
Specifically, the selected subcarrier and sensing symbol indexes for the three subcarrier distribution schemes to be evaluated are listed in Table \ref{tab:3}. As depicted in Fig. \ref{figure5} and Fig. \ref{figure6}, we observe that the edge-first scheme outperforms the interleaved scheme, while the interleaved scheme surpasses the subband scheme in terms of both CRB and MSE. This phenomenon arises because the variance of the subcarrier indexes is largest in the edge-first scheme, larger in the interleaved scheme, and smallest in the subband scheme. 
\begin{table}[h]
	\scriptsize
	\caption{Subcarrier/sensing symbol distribution schemes in single-UE OFDMA PMNs.}
	\label{tab:3}       
	\centering
	\begin{tabular}{l|c} 
		\hline 
		Scheme&Subcarrier index \\
		\hline
		Subband distribution & [1 2 3 4 5 6 7 8 9 10 11 12 13 14 15 16]\\
		\hline
		Interleaved distribution& [1 4 7 10 13 16 19 22 25 28 31 34 37 40 43 46]\\
		\hline
		Edge-first distribution& [1 2 3 4 5 6 7 8 41 42 43 44 45 46 47 48]\\
		\hline
	\end{tabular}%
\end{table}%
\begin{table}[t]
	\scriptsize
	\caption{Subcarrier distribution schemes in a three-UE OFDMA PMN.}
	\label{tab:2}       
	\centering
	\begin{tabular}{l|c} 
		\hline 
		Scheme&Subcarrier index \\
		\hline
		Generalized  &UE 1:\ [1 2 5 8 9 13 16 17 19 25 26 29 34 36 38 41]\\  distribution
		&UE 2:\ [3 6 7 11 14 15 20 23 27 31 32 35 37 43 46 48]\\  
		&UE 3:\ [4 10 12 18 21 22 24 28 30 33 39 40 42 44 45 47]\\
		\hline
		Interleaved&UE 1:\ [1 4 7 10 13 16 19 22 25 28 31 34 37 40 43 46]\\ distribution
		&UE 2:\ [2 5 8 11 14 17 20 23 26 29 32 35 38 41 44 47]\\ 
		&UE 3:\ [3 6 9 12 15 18 21 24 27 30 33 36 39 42 45 48]\\
		\hline
		Edge-first&UE 1:\ [1 2 3 4 5 6 7 8 41 42 43 44 45 46 47 48]\\ distribution
		&UE 2:\ [9 10 11 12 13 14 15 16 33 34 35 36 37 38 39 40]\\ 
		&UE 3:\ [17 18 19 20 21 22 23 24 25 26 27 28 29 30 31 32]\\
		\hline
	\end{tabular}%
\end{table}%
\begin{figure}[t]
	\centering
	\hspace{-6pt}
	\includegraphics[width=2.9in]{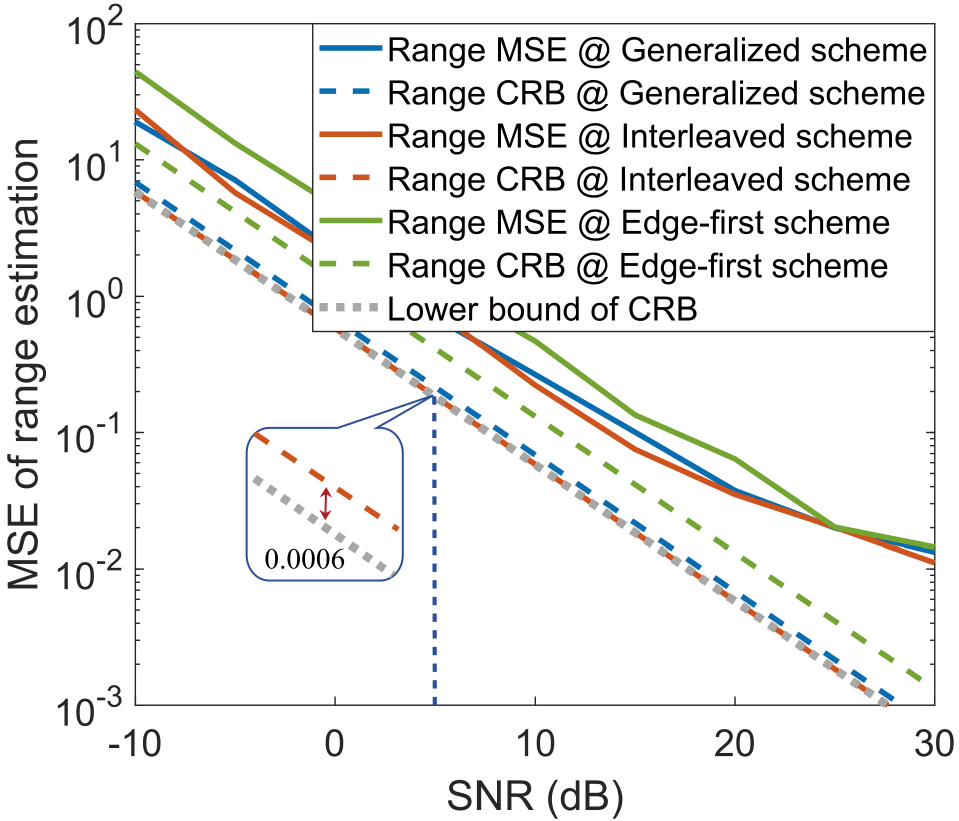}
	\caption{The maximum range estimation MSE and CRB over multiple UEs with different subcarrier distribution in multi-UE OFDMA PMNs.}\label{figure8}
\end{figure}

In multi-UE OFDMA PMNs, the maximum MSE and CRB over multiple UEs under different subcarrier distribution schemes are illustrated in Fig. \ref{figure8}. In the simulation, we simulate three UEs, each of which is assigned 16 subcarriers out of a total of 48 subcarriers. The subcarrier indexes for the three schemes to be evaluated are listed in Table \ref{tab:2}. As presented in Fig. \ref{figure8}, the interleaved subcarrier distribution outperforms the generalized distribution and the edge-first distribution, while the generalized distribution surpasses the edge-first distribution. Additionally, the lower bound of the maximum CRB over multiple UEs is included in the figure. The figure suggests that the CRB gap  $[{\textrm{CRB}_\textrm{I}(R)-\textrm{CRB}_\textrm{low}(R)}]/{\operatorname{CRB}_\textrm{low}(R)}$ is negligible, aligning with the theoretical analysis in Proposition \ref{pro6}. For practical system design, this insight suggests that the interleaved scheme is sufficiently effective in minimizing the maximum CRB of multiple UEs. The marginal CRB gap may not justify the significant computational cost required to search for the optimal distribution scheme.

\begin{figure}[t]
	\centering
	\hspace{6pt}
	\includegraphics[width=2.9in]{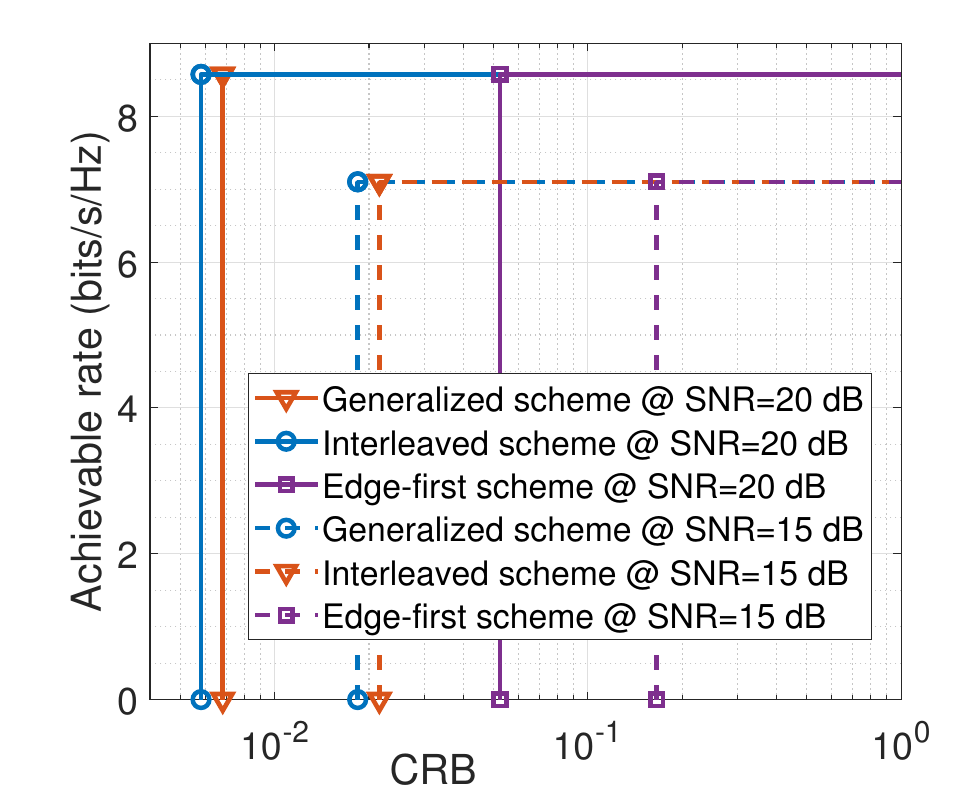}
	\caption{The maximum CRB over multiple UEs and the achievable communication rates with different subcarrier distribution in a three-UE OFDMA PMN. The subcarrier distribution schemes for the three UEs are listed in Table. \ref{tab:2}.}\label{figure10}
\end{figure}

In Fig. \ref{figure10}, the maximum CRB over multiple UEs and the achievable communication rates with different subcarrier distribution schemes are evaluated in a three-UE OFDMA PMNs. In the simulation, the ranges and velocities of the targets for each UE are set to [30, 50, 80] meters and [10, 30, 20] meters per second, respectively.
As shown in Fig. \ref{figure10}, the achievable communication rates remain relatively constant across different subcarrier distribution schemes, while the maximum CRB over multiple UEs varies significantly. Specifically, transforming the generalized subcarrier distribution into the interleaved subcarrier distribution yields an improvement in the maximum CRB over multiple UEs. Additionally, there is a significant reduction in the maximum CRB over multiple UEs when the distribution shifts from the generalized scheme to the edge-first scheme. The results imply that optimizing the subcarrier distribution can potentially enhance the sensing performance of the entire system while maintaining constant communication performance.

\section{Conclusion}

The subcarrier and sensing symbol distribution in OFDMA PMNs are investigated aiming to enhance sensing and communication capabilities. First, we derived the closed-form expression for the range and velocity CRB under general subcarrier and sensing symbol distributions. 
The expression revealed that the range and velocity CRB are inversely proportional to the variances of the subcarrier and sensing symbol indexes, respectively. Based on the CRB expressions, we identified the edge-first scheme as the best and the subband scheme as the worst for both subcarrier and sensing symbol distribution in single-UE OFDMA PMNs. Moreover, in multi-UE OFDMA PMNs, we proposed a tight lower bound for the minimized maximum CRB of multiple UEs, and proved that the gap between the lower bound and the CRB under the interleaved subcarrier distribution is negligible. Furthermore, our theoretical proof also confirmed that the impact of subcarrier distribution on achievable communication rates in narrowband synchronous systems is marginal. Numerical simulations were conducted to validate our theoretical analysis.

\appendices
\section{Proof of the proposition 1}
	In the following, we complete the proof in two steps.

	\textit{Step 1}: The first equation in (\ref{condition}) aims to cancel the data information of the $k$th UE. Based on this equation, ${\bf C}_k$ for $k=1,\cdots,K$ should satisfy
	\begin{equation}\label{equ35}
		\begin{cases}
			&{\bf C}_k[{\boldsymbol{\zeta}}_k[n],{\boldsymbol{\zeta}}_k[n]]\!=({\bf D}_{g,k}[{\boldsymbol{\zeta}}_k[n],{\boldsymbol{\zeta}}_k[n]])^{-1}, \\
			&{\bf C}_k[{\boldsymbol{\zeta}}_k[n],\!i]\!=\!0,i=1\!,\!\cdots\!,\!N, \textrm{and}\ i\neq {\boldsymbol{\zeta}}_k[n],\\
		\end{cases}
	\end{equation}
	where $n=1,\cdots,N_k$. Actually, (\ref{equ35}) implies that there are many feasible ${\bf C}_{k}$ satisfying the first equation in (\ref{condition}). 

\textit{Step 2}: The second equation in (\ref{condition}) aims to nullify interference caused by subcarriers belonging to other UEs during the compensation. 
Intuitively, if subcarriers are fully assigned to UE and the subcarriers belonging to different UEs do not overlap, the $i$th column, $i\in\{\boldsymbol{\zeta}_{k^{'}}[n],n=1,\cdots,N_k\}$, of ${\bf C}_k$ should be 0 to achieve ${\bf C}_k{\bf D}_{g,k^{'}}={\bf 0}_{N\times N}$ for any $k^{'}\neq k$. Therefore, ${\bf C}_k$ have to be satisfied
\begin{equation}\label{condition2}
		{\bf c}_k[n]=
		\begin{cases}
			({\bf D}_{g,k}[n,n])^{-1},&\ {\bf D}_{g,k}[n,n]\neq 0,\\
			0,&\ {\bf D}_{g,k}[n,n]= 0,\\
		\end{cases}
\end{equation}
where ${\bf C}_k={\textrm{diag}}({\bf c}_k)$. This thus gives us the feasible ${\bf C}_k$ for scenarios where subcarriers do not overlap.

For scenarios where some subcarriers are multiplexed by different UEs, 
we reformulate the second equation in (\ref{condition}) as
\begin{equation}\label{condition1}
	\Big(\sum\nolimits_{k^{'}\neq k}^{K}{{\bf B}_{k^{'}}}{\bf D}_{g,k^{'}}\Big){\bf C}_{k} = {\bf 0}_{N\times N},
\end{equation}
where ${\bf B}_{k^{'}}$ is denoted as $\sum_{l=1}^{L_{k^{'}}}\tilde\alpha_{k^{'},l}{\bf a}(\Omega_{k^{'},l}^{\textrm{r}}){\boldsymbol\tau}_{k^{'},l}$. Moreover, ${\bf B}_{k^{'}}$ can be expressed as $[{\bf b}_{k^{'}}^1,\cdots,{\bf b}_{k^{'}}^n,\cdots,{\bf b}_{k^{'}}^N]$, where ${\bf b}_{k^{'}}^n\in\mathbb{C}^{M_\textrm{R}\times1}$ is the $n$th column of ${\bf B}_{k^{'}}$. Then, we denote the diagonal matrix ${\bf \tilde{D}}_{g,k^{'},k}$ as ${\bf D}_{g,k^{'}}{\bf C}_{k}$ for brevity. 
Consequently, for any ${k^{'}}\neq k$, (\ref{condition1}) can be further expressed as
\begin{equation}
		[{\bf B}_1,\cdots,{\bf B}_K]\cdot[
		{\bf \tilde{D}}_{g,1,k}^{\textrm{T}},\cdots,{\bf \tilde{D}}_{g,K,k}^{\textrm{T}}
		]^{\textrm{T}}={\bf 0}_{N\times N}.
\end{equation}%
Furthermore, we denote the indices of UE sharing the $n$th subcarrier with the $k$th UE as the set ${\mathbb{U}_n}$. 
Then we obtain
\begin{fequation}
		\Big[{\bf c}_{k}[1]\!\sum\limits_{{\!\!k^{'}}\in{\mathbb U}_1}\!{\bf b}_{k^{'}}^1{\bf s}_{g,k^{'}}[1],\cdots,{\bf c}_{k}[N]\!\sum\limits_{{\!\!k^{'}}\in{\mathbb U}_N}\!{\bf b}_{k^{'}}^N{\bf s}_{g,k^{'}}[N]\Big]\!={\bf 0}_{N \times N}.
		\label{simple}
\end{fequation}%
Since ${\bf b}_{k^{'}\!}^n$ is a linear combination of $L_{k^{'}\!
}$ receiving steering vectors ${\bf a}{(\Omega_{k^{'}\!, l}^{\textrm{r}})}$, $ l\!=\!1,\!\cdots\!,L_{k^{'}}$, with random AOAs, $\{\mathbf{b}_{k^{'}}^n\}_{k^{'} \in\mathbb{U}_n}$ are hardly linearly correlated. Thus, $\!\sum_{{k^{'}}\in{\mathbb U}_n}{\bf b}_{k^{'}}^1{\bf s}_{g,k^{'}}[n]$ can hardly be zero. Moreover, since ${\bf c}_{k}[n]$ for $n=1,\cdots,N$ cannot all be zero, there is no ${\bf C}_{k}$ satisfying (\ref{simple}).

As a result, to successfully conduct data compensation in OFDMA systems, each subcarrier can only be allocated to at most one UE. In such cases, the compensation matrix ${\bf C}_k$ can be designated as (\ref{condition2}).

\section{Proof of the Proposition \ref{pro1_1}}\label{Pro3}
In this proof, we demonstrate that the variance of the subcarrier indexes of the edge-first scheme is constantly larger than that of any other schemes.

Specifically, we start by denoting the subcarrier index vectors under the edge-first distribution scheme and one other distribution scheme as ${\boldsymbol{\zeta}}_{k,{\textrm{O}}}$ and ${\boldsymbol{\zeta}}_{k,{\textrm{A}}}$, respectively.
Then, we intend to verify whether ${\boldsymbol{\zeta}}_{k,{\textrm{O}}}$ and ${\boldsymbol{\zeta}}_{k,{\textrm{A}}}$ constantly satisfy
\begin{sequation}\label{equ44}
	\begin{aligned}
	&\Big\{{1}/{N_k}\sum\nolimits_{n=1}^{N_k}\boldsymbol{\zeta}_{k,{\textrm{O}}}[n]^2-{1}/{N_k^2}\left[\sum\nolimits_{n=1}^{N_k}\boldsymbol{\zeta}_{k,{\textrm{O}}}[n]\right]^2\Big\}-\\
	&\Big\{{1}/{N_k}\sum\nolimits_{n=1}^{N_k}\boldsymbol{\zeta}_{k,{\textrm{A}}}[n]^2-{1}/{N_k^2}\left[\sum\nolimits_{n=1}^{N_k}\boldsymbol{\zeta}_{k,{\textrm{A}}}[n]\right]^2\Big\}>0.
	\end{aligned}
\end{sequation}%
Furthermore,  (\ref{equ44}) can be reformulated as
\begin{fequation}\label{equ59}
	\begin{aligned}{}
		&{1}/{N_k}\sum\nolimits_{n=1}^{N_k}(\boldsymbol{\zeta}_{k,{\textrm{O}}}[n]+\boldsymbol{\zeta}_{k,{\textrm{A}}}[n])(\boldsymbol{\zeta}_{k,{\textrm{O}}}[n]-\boldsymbol{\zeta}_{k,{\textrm{A}}}[n])-{1}/{N_k^2}\\
		&\sum\nolimits_{n=1}^{N_k}(\boldsymbol{\zeta}_{k,{\textrm{O}}}[n]+\boldsymbol{\zeta}_{k,{\textrm{A}}}[n])\sum\nolimits_{n_k=1}^{N_k}(\boldsymbol{\zeta}_{k,{\textrm{O}}}[n]-\boldsymbol{\zeta}_{k,{\textrm{A}}}[n])>0.
	\end{aligned}
\end{fequation}%
For brevity, let us denote $\boldsymbol{\zeta}_{k,{\textrm{O}}}[n]+\boldsymbol{\zeta}_{k,{\textrm{A}}}[n]$, $\boldsymbol{\zeta}_{k,{\textrm{O}}}[n]-\boldsymbol{\zeta}_{k,{\textrm{A}}}[n]$, ${1}/{N_k}\sum_{n=1}^{N_k}(\boldsymbol{\zeta}_{k,{\textrm{O}}}[n]+\boldsymbol{\zeta}_{k,{\textrm{A}}}[n])$, and ${1}/{N_k}\sum_{n=1}^{N_k}(\boldsymbol{\zeta}_{k,{\textrm{O}}}[n]-\boldsymbol{\zeta}_{k,{\textrm{A}}}[n])$ as $\boldsymbol{\mu}_k[n]$, $\boldsymbol{\rho}_k[n]$, ${\bar{\mu}}_k$, and $\bar\rho_k$, respectively. Thus, (\ref{equ59}) can be represented as
\begin{equation}\label{equ60}
	{1}/{N_k}\sum\nolimits_{n=1}^{N_k}{\boldsymbol\mu}_k[n]\boldsymbol{\rho}_k[n]-{\bar{\mu}}_k\bar\rho_k>0.
\end{equation}

\begin{figure}[tbp]
	\centering
	\includegraphics[width=2.7in]{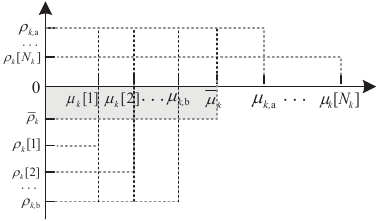}
	\caption{Principle of proof $\frac{1}{N_k}\sum_{n=1}^{N_k}{\boldsymbol\mu}_k[n]\boldsymbol{\rho}_k[n]-{\bar{\mu}}_k\bar\rho_k\!>\!0$. According to definition of $\boldsymbol{\zeta}_{k,{\textrm{O}}}$ and $\boldsymbol{\zeta}_{k,{\textrm {A}}}$, there establish $\bar{\rho}\le0,\rho_{k,{\textrm a}}\le\!\cdots\!\le\boldsymbol{\rho}_k[1]\le0$, and $\rho_{k,{\textrm a}}\ge\!\cdots\!\ge\!\boldsymbol{\rho}_k[N_k] \! \ge0$. $\rho_{k,{\textrm a}}$ and $\rho_{k,{\textrm b}}$ represent the first element before and after $\bar{\rho}_k$. $\mu_{k,{\textrm a}}$ and $\mu_{k,{\textrm b}}$ represent the first element before and after $\bar{\mu}_k$.
	\label{Pro_3}}
	\vspace*{-12pt}
\end{figure}

In the following, we provide a proof of (\ref{equ60}), the principle of which is illustrated in Fig. \ref{Pro_3}. Without loss of generality, we assume $\bar{\rho}_k=\frac{1}{N_k}\sum_{n=1}^{N_k}(\boldsymbol{\zeta}_{k,{\textrm {O}}}[n]-\boldsymbol{\zeta}_{k,{\textrm{A}}}[n])\le0$ in the proof. In fact, $\bar{\rho}_k$ can be either greater than, equal to, or less than $0$ under actual subcarrier distribution schemes. However, for any $\boldsymbol{\zeta}_{k,{\textrm{O}}}$ and $\boldsymbol{\zeta}_{k,{\textrm {A}}}$, whose means are defined as $\bar{\rho}_{k,{\textrm{O}}}$ and $\bar{\rho}_{k,{\textrm{A}}}$, satisfy $\bar{\rho}_k=\bar{\rho}_{k,{\textrm{O}}}-\bar{\rho}_{k,{\textrm{A}}}\ge0$, namely $\bar{\rho}_k\ge0$, there must be a corresponding symmetric scheme with means $\sum_{n=1}^{N}n-\bar{\rho}_{k,{\textrm{O}}}$ and $\sum_{n=1}^{N}n-\bar{\rho}_{k,{\textrm {A}}}$. 
For instance, in a 10-subcarrier system, given $\boldsymbol{\zeta}_{k,{\textrm{O}}}=[1,2,10]$ and $\boldsymbol{\zeta}_{k,{\textrm{A}}}=[1,2,5]$, there symmetric scheme are $[10,9,1]$ and $[10,9,6]$, respectively. Obviously, their symmetric schemes satisfy $\bar{\rho}_k<0$. Moreover, it is important to note that the variances of the original and symmetric schemes are the same. Therefore, if the symmetric schemes satisfying $\bar{\rho}_k\le0$ meet condition (\ref{equ60}), the original schemes satisfying $\bar{\rho}_k\ge0$ will also satisfy (\ref{equ60}). Thus, the assumption $\bar{\rho}_k\le0$ is reasonable and will not limit the scope of the conclusion. 

In this proof, we represent  ${\bar{\mu}}_k\bar\rho_k$ and ${\boldsymbol\mu}_k[n]\boldsymbol{\rho}_k[n], n\!=\!1,\!\cdots\!,N_k$ by the rectangular areas in Fig. \ref{Pro_3}. Therefore, proving ${1}/{N_k}\sum_{n=1}^{N_k}{\boldsymbol\mu}_k[n]\boldsymbol{\rho}_k[n]-{\bar{\mu}}_k\bar\rho_k>0$ is equivalent to demonstrating that the mean of the white rectangular areas ${\boldsymbol\mu}_k[n]\boldsymbol{\rho}_k[n]$, $n=1,\!\cdots\!,N_k$ is greater than the area of the grey rectangle ${\bar{\mu}}_k\bar\rho_k$. 
The specific proof is as follows.

\textit{Step 1:} For the rectangles in the first column, there holds
\begin{equation}
	\boldsymbol{\mu}_k[1](\sum\nolimits_{n=1}^{N_k}{\boldsymbol\rho}_k[n]-\bar{\rho}_k)=0.
\end{equation}

\textit{Step 2:} For the rectangle in the second column, since $\boldsymbol{\rho}_k[1]<0$, there holds
\begin{fequation}
	(\boldsymbol{\mu}_k[2]\!-\boldsymbol{\mu}_k[1])(\sum\nolimits_{n=2}^{N_k}\!\boldsymbol{\rho}_k[n]-\bar{\rho}_k)\!=\!-(\boldsymbol{\mu}_k[2]-\boldsymbol{\mu}_k[1])\boldsymbol{\rho}_k[1]\!>\!0.
\end{fequation}

\textit{Step p ($p\le b+1$):} For the rectangle in the $p$th column, there holds
\begin{fequation}
	\begin{aligned}
		(\boldsymbol{\mu}_k[p]\!-\!\boldsymbol{\mu}_k[p-1])(\sum\limits_{n=p}^{N_k}\!\boldsymbol{\rho}_k[n]-\bar{\rho}_k)\!=\!-(\boldsymbol{\mu}_k[2]\!-\!\boldsymbol{\mu}_k[1])\!\sum\limits_{n=1}^{p-1}\!\!\boldsymbol{\rho}_k[n]\!>\!0.
	\end{aligned}
\end{fequation}

\textit{Step q ($ a\le q\le N_k+1$):} For the rectangle in the $q$th column, it holds
\begin{equation}
	(\boldsymbol{\mu}_k[q]\!-\!\boldsymbol{\mu}_k[q-1])(\sum\nolimits_{n=q}^{N_k}\boldsymbol{\rho}_k[n])>0.
\end{equation}

By summing all equations from step $1$ to $N_k+1$, we obtain:
\begin{equation}
	{1}/{N_k}\sum\nolimits_{n=1}^{N_k}{\boldsymbol\mu}_k[n]\boldsymbol{\rho}_k[n]-{\bar{\mu}}_k\bar\rho_k>0.
\end{equation}

\section{Proof of the Proposition 3}\label{Pro2}
We represent $\boldsymbol\zeta_{k,{\textrm{S}}}$ as the subcarrier index vector of the $k$th UE when using the subband distribution scheme. For $ n=1,\!\cdots\!,N_k-1$, we assume $\boldsymbol\zeta_{k,{\textrm {S}}}[n+1]>\boldsymbol\zeta_{k,{\textrm{S}}}[n]$. In addition, for simplicity, we will assume that the minimum value in $\boldsymbol\zeta_{k,{\textrm{S}}}$, i.e. $\boldsymbol\zeta_{k,{\textrm{S}}}[1]$, is $0$. This assumption does not affect the following analysis. 
Similarly, we define ${\boldsymbol\zeta}_{k,{\textrm{A}}}$ as the subcarrier index vector of the $k$th UE when using any other distribution scheme and ${\boldsymbol\zeta}_{k,{\textrm{A}}}=\boldsymbol\zeta_{k,{\textrm{S}}}+\boldsymbol{\rho}_k$. It is evident that $\boldsymbol{\rho}_k[N_k]\geq\boldsymbol{\rho}_k[n]\geq\boldsymbol{\rho}_k[1]=0$ for $n=2,\cdots,N_k-1$, and $\boldsymbol{\rho}_k[n]$, for $n=2,\cdots,N_k$ are not all equal to zero. 

Obviously, if the variance of any ${\boldsymbol\zeta}_{k,{\textrm{A}}}$ is constantly larger than that of $\boldsymbol\zeta_{k,{\textrm{S}}}$, Proposition $2$ is proved. To achieve the proof, we first denote the variance of  ${\boldsymbol\zeta}_{k,{\textrm{A}}}$ and $\boldsymbol\zeta_{k,{\textrm{S}}}$ as $\tilde{A}$ and $A$, respectively, and formulate $\tilde{A}$ and $A$ as
\begin{equation}
		\tilde{A}={1}/{N_k}\sum\nolimits_{n=1}^{N_k}{\boldsymbol\zeta}_{k,{\textrm {A}}}^2[n]-[\frac{1}{N_k}\sum\nolimits_{n=1}^{N_k}{\boldsymbol\zeta}_{k,{\textrm {A}}}[n]]^2,
	\end{equation}
\begin{equation}
		A={1}/{N_k}\sum\nolimits_{n=1}^{N_k}\boldsymbol\zeta_{k,{\textrm {S}}}^2[n]-[\frac{1}{N_k}\sum\nolimits_{n=1}^{N_k}\boldsymbol\zeta_{k,{\textrm{S}}}[n]]^2.
\end{equation}%
Furthermore, $\tilde{A}-A\geq0$ can be formulated as
\begin{fequation}\label{compare}
	\begin{aligned}
	\tilde{A}-A
	&=\!2(\frac{1}{N_k}\sum_{n=1}^{N_k}\boldsymbol\zeta_{k,{\textrm {S}}}[n]\boldsymbol{\rho}_k[n]\!-\!\bar\rho_k\bar{\zeta}_k)\!+\!\frac{1}{N_k}\sum_{n=1}^{N_k}\boldsymbol{\rho}_k[n]^2\!-\!\bar\rho_k^2,
	\end{aligned}
\end{fequation}%
where $\bar\rho_k={1}/{N_k}\sum_{n=1}^{N_k}\boldsymbol{\rho}_k[n]$, and $\bar{\zeta}_k/{N_k}=\sum_{n=1}^{N_k}\boldsymbol\zeta_{k,{\textrm{S}}}[n]$, respectively. It is evident that ${1}/{N_k}\sum_{n=1}^{N_k}\boldsymbol{\rho}_k[n]^2-\bar\rho_k^2$ is the variance of elements in $\boldsymbol{\rho}_k$, and it is always larger than zero. Hence, if the inequality, ${1}/{N_k}\sum_{n=1}^{N_k}\boldsymbol\zeta_{k,{\textrm{S}}}[n]\boldsymbol{\rho}_k[n]-\bar\rho_k\bar{\zeta}_k\geq0$ always holds, the proposition is proven.

Furthermore, we reformulate ${1}/{N_k}\sum_{n=1}^{N_k}\boldsymbol\zeta_{k,{\textrm{S}}}[n]\boldsymbol{\rho}_k[n]-\bar\rho_k\bar{\zeta}_k\geq0$ as
\begin{sequation}
			\label{equ55}
			\sum\nolimits_{n=1}^{N_k}\boldsymbol\zeta_{k,{\textrm{S}}}[n]\boldsymbol{\rho}_k[n]\!-\!\frac{1}{N_k}\sum\nolimits_{n=1}^{N_k}\boldsymbol{\rho}_k[n]\sum\nolimits_{n\!-\!1}^{N_k}\boldsymbol\zeta_{k,{\textrm{S}}}[n]\!\geq\!0,
\end{sequation}%
which takes the form of the rearrangement inequality \cite{day1972rearrangement}. Therefore, inequality (\ref{equ55}) evidently establishes, proving  $\tilde{A}-A\geq0$. 
This completes the proof that the sub-band distribution results in the least variance and consequently the worst CRB.

\section{Proof of the proposition 4}\label{proof4}

According to \cite{scheffe1999analysis}, the relationship between the variance of a whole set, the variance within the subsets, and the variance between subsets satisfies
\begin{equation}
	\label{equ66}
	\sum\limits_{n=1}^{N}\!(n-\bar{n})^2\!\!=\!\sum\limits_{k=1}^{K}\sum\limits_{i=1}^{N_k}(n_{k,i}-\bar{n}_k)^2+\sum\limits_{k=1}^{K}\!N_k(\bar{n}_k-\bar{n})^2,
\end{equation}%
where $\bar{n}$, $n_{k,i}$, and $\bar{n}_k$ are the means of all subcarrier indexes, the index of the $i$th subcarrier of the $k$th UE, and the mean of subcarrier indexes of the $k$th UE, respectively. 
Thus, it is evident that $\bar{n}={1}/{N}\sum_{n=1}^{N}n$ and $\bar{n}_k=1/N_k\sum_{i=1}^{N_k}n_{k,i}$, and when the total number of subcarriers is determined, $\sum_{n=1}^{N}(n-\bar{n})^2$ is a constant.

Moreover, since $\sum_{k=1}^{K}N_k(\bar{n}_k-\bar{n})^2$ must be not less than 0, the sum of the variance within the subset satisfies
\begin{equation}
	\sum\nolimits_{k=1}^{K}\sum\nolimits_{i=1}^{N_k}(n_{k,i}-\bar{n}_k)^2\le\sum\nolimits_{n=1}^{N}(n-\bar{n})^2.
\end{equation}
We then assume that the maximized minimum variance is achieved by the $k_{\textrm{m}}$th UE and denote the variance as $\epsilon$. It is obvious that  $\epsilon$ must satisfy
\begin{sequation}\label{equ69}
	N_{k_{\textrm{m}}}\epsilon\le\!\!\frac{1}{K}\sum\nolimits_{k=1}^{K}\!\sum\nolimits_{i=1}^{N_k}\!(n_{k,i}\!-\bar{n}_k)^2\le\frac{1}{K}\sum\nolimits_{n=1}^{N}(n-\bar{n})^2.
\end{sequation}%
It is intuitive that the equality in (\ref{equ69}) holds if and only if the mean and variance of different subcarrier index subsets are the same. 
This thus completes the proof.

\section{Proof of the proposition \ref{pro5}}\label{proof5}
According to the constraint (\ref{equ32}), it is evident that the $K$ index subsets have the same variances. Moreover, the means of the $K$ index subsets are $\bar{n}_1, \bar{n}_1+\eta_{1,2},\cdots,\bar{n}_1+\eta_{1,K}$. Therefore, we have
\begin{equation}\label{equ64}
	KN_k\epsilon=\sum\nolimits_{n=1}^{N}(n-\bar{n})^2-\sum\nolimits_{k=1}^{K}N_k(\bar{n}_k-\bar{n})^2.
\end{equation}
Since $\sum_{n=1}^{N}(n-\bar{n})^2$ is a constant, to maximize $\epsilon$, the term $\sum_{k=1}^{K}N_k(\bar{n}_k-\bar{n})^2$ should be minimized.

With the conclusion in Proposition \ref{pro1},  $\sum_{k=1}^{K}N_k(\bar{n}_k-\bar{n})^2=N_k\sum_{k=1}^{K}(\bar{n}_k-\bar{n})^2$ is minimized if the indices $\bar{n}_1, \bar{n}_1+\eta_{1,2},\cdots,\bar{n}_1+\eta_{1,K}$ follow a subband distribution, i.e.  $\eta_{1,k}=k-1$. Obviously, the subcarrier indexes of the interleaved distribution scheme satisfy this condition.

This thus completes the proof of the Proposition \ref{pro5}.
\section{Proof of the proposition \ref{pro6}}\label{proof6}

Based on (\ref{equ27}), (\ref{equ34_1}), and (\ref{equ64}), we formulate the gap $V=\frac{\textrm{CRB}_\textrm{I}(R)-\textrm{CRB}_\textrm{low}(R)}{\textrm{CRB}_\textrm{low}(R)}$ as
\begin{equation}\label{equ72}
	\begin{small}
		\begin{aligned}
			V=&\frac{\sum_{n=1}^{N}(n-\bar{n})^2}{\sum_{n=1}^{N}(n-\bar{n})^2-\sum_{k=1}^{K}N_k(\bar{n}_k-\bar{n})^2}-1.
		\end{aligned}
	\end{small}
\end{equation} 

Depending on the parity of $K$, the summation process of $\sum_{k=1}^{K}N_k(\bar{n}_k-\bar{n})^2$ varies. Therefore, we discuss the value of $\sum_{k=1}^{K}N_k(\bar{n}_k-\bar{n})^2$ in two categories. 
\begin{fequation}
	\begin{cases}
		\begin{aligned}
			2N_k[1^2+\!\cdots+\!({K\!-\!1}/{2})^2]\! =\!{(K\!-\!1)(K\!+\!1)N}/{12}, \ & K\  \textrm{is odd}\\
			2N_k[({1}/{2})^2+\!\cdots\!+({K-1}/{2})^2]=\!{(K\!-\!1)(K\!+\!1)N}/{12}, \ & K\  \textrm{is even}.\\
		\end{aligned}
	\end{cases}
\end{fequation}
Similarly, $\sum_{n=1}^{N}(n-\bar{n})^2$ can be represented as
\begin{fequation}
	\begin{cases}
		\begin{aligned}
			2[1^2\!+\!\cdots\!\!+\!({N\!-\!1}/{2})^2]\!\!&=\!{N(N\!-\!1)(N\!+\!1)}/{12}, N\ \!  \textrm{is odd,}\!\!\\
			2[({1}/{2})^2\!+\!\cdots\!+\!({N\!-\!1}/{2})^2]\!\!&=\!{N(N\!-\!1)(N\!+\!1)}/{12}, N\ \! \textrm{is even.}\!\!\\
		\end{aligned}
	\end{cases}
\end{fequation}As a result, (\ref{equ72}) can be simplified as
\begin{equation}\label{equ76}
	\begin{small}\begin{aligned}
			{V}=\frac{{(K-1)(K+1)}}{{(N-1)(N+1)}-{(K-1)(K+1)}}.
\end{aligned}\end{small}\end{equation} This completes the proof.



\ifCLASSOPTIONcaptionsoff
\newpage
\fi



\bibliographystyle{IEEEtran}
\bibliography{IEEEabrv,reference.bib}

\end{document}